\documentclass[aps,prx,notitlepage,superscriptaddress,showpacs,twocolumn ]{revtex4-1}
\usepackage{graphicx,subfigure,epsfig}
 \usepackage{array,multirow}
\usepackage{dcolumn}
\usepackage{amssymb,amsmath,amsfonts,mathrsfs}
\usepackage{array}
\usepackage{times,setspace}
\usepackage{latexsym}
\usepackage{float,flafter,bm,bbm}
\usepackage{epstopdf,color,multirow}
\usepackage[colorlinks,linkcolor=blue,anchorcolor=blue,urlcolor=blue,citecolor=blue]{hyperref}
\usepackage{footnote}

\hypersetup{
    colorlinks=true,
    linkcolor=blue,
    filecolor=magenta,
    urlcolor=blue,
}

\begin{document}

\title{Doped Mott Insulators in the Triangular Lattice Hubbard Model}
\author{Zheng Zhu}
\affiliation{Department of Physics, Harvard University, Cambridge, Massachusetts 02138, USA}
\affiliation{Kavli Institute for Theoretical Sciences, University of Chinese Academy of Sciences, Beijing 100190, China}
\author{D. N.  Sheng}
 \affiliation{Department of Physics and Astronomy, California State University, Northridge, California 91330, USA}
\author{Ashvin Vishwanath}
\affiliation{Department of Physics, Harvard University, Cambridge, Massachusetts 02138, USA}
 \begin{abstract}
We investigate the evolution of the Mott insulators in the triangular lattice Hubbard Model,  as a function of hole doping $\delta$ in both the strong  and intermediate coupling  limits.  Using the advanced density matrix renormalization group (DMRG) method,  at light  hole doping  $\delta\lesssim 10\%$, we find  a significant difference between strong and intermediate couplings.    Notably, at intermediate coupling an unusual metallic state emerges, with  short ranged spin correlations but long ranged spin-chirality order. Moreover, no clear Fermi surface or wave-vector is observed, this chiral metal also exhibits staggered  loop current, which breaks the translational symmetry. These features disappear on increasing interaction strength or on further doping. At strong  coupling,  the 120 degree magnetic order of the insulating magnet persists for light doping, and  produces hole pockets with a well defined Fermi surface. On further doping,  $\delta \approx 10\%\sim 20\%$  SDW order and coherent hole Fermi pockets are found at both strong and intermediate couplings.  At even higher doping $\delta \gtrsim 20\%$, the SDW order is suppressed and the spin-singlet Cooper pair correlations are  simultaneously enhanced.
We also briefly comment on the strong particle-hole asymmetry of the model.
 \end{abstract}
\date{\today}
\maketitle

\section{Introduction}

A central issue in the physics of strongly correlated materials is the nature of the correlated phases that emerge on doping a Mott insulator. Given its  relevance to  the high temperature cuprate superconductors, much effort have been devoted to doped Mott insulators in the square lattice Hubbard Model~\cite{PatrickLeeRMP,Anderson,Dagotto1994,Zaanen2015}. However, the analogous problem on the triangular lattice is equally interesting and likely to exhibit new and distinct physics, due to magnetic frustration and the absence of nesting and particle-hole symmetry  in the minimal models.

Experimentally, the discovery of the spin liquids~\cite{QSL1,QSL2} in organic materials~\cite{shimizu,itou1, itou2, yamashita1, yamashita2,isono}  like $\kappa$-(BEDT-TTF)$_2$Cu$_2$(CN)$_3$, EtMe$_3$Sb[Pd(dmit)$_2$]$_2$ and $\kappa$-H$_3$(Cat-EDT-TTF)$_2$ has triggered substantial efforts on the triangular lattice Hubbard model, which is suggested to be the simplest model to understand unconventional correlated physics in these systems.  More recently, the triangular lattice magnets including Ba$_3$CuSb$_2$O$_9$\cite{BaCuSb},   Yb$_2$Ti$_2$O$_7$\cite{Balents2011}, Cs$_2$CuCl$_4$ \cite{CsCuCl1,CsCuCl2}, Ba$_3$CoSb$_2$O$_9$\cite{BatistaBaCoSb},  YbMgGaO$_4$\cite{YbMgGaO},  TbInO$_3$\cite{TbInO}, NaYbO$_2$\cite{NaYbO1,NaYbO2}, the transition-metal dichalcogenide (TMD)~\cite{PatrickLee2017,Macdonald2018} as well as twisted bilayers of TMD ~\cite{Macdonald2019,TwistedTMD,TwistedTMD1} have been successively suggested to realize triangular lattice Hubbard models or their effective Heisenberg models. In a different background,
the triangular Hubbard model was recently  realized on optical lattices\cite{Yang2021} with loading ultracold fermions~\cite{ColdAtom1,ColdAtom2, ColdAtom3}. In these platforms, both the coupling strength $U/t$ and charge concentration are widely tunable and accurately controllable, which allow for probing the correlated electron physics on frustrated lattices.

Theoretically, the triangular lattice Hubbard model still  poses a great challenge.  At half filling,  mean field approaches~\cite{MF1,MF2,Sachdev1992,Wang2006, Song2019} and numerical studies~\cite{Huse1988,Motrunich2005,Koretsune2007,Watanabe2008,Phillips2009,Tocchio2013,Sahebsara2008,Clay2008,Yang2010,Yoshioka2009,Kokalj2013,Li2014,Yamada2014,Laubach2015,Mishmash2015,Iqbal2016,Shirakawa2017,Szasz2018,Hu2019,Szasz2021,Wietek2021,Chen2021} have identified the metal-insulator transition (MIT) around $U_c/t\approx$5$\sim$8, and two distinct Mott insulators: a $120^{\circ}$ N\'eel ordered phase in the strong coupling $U\gg t$ limit, and a quantum disordered state, potentially a spin liquid  phase, around the MIT.
Away from half-filling, the slave-boson mean field~\cite{SBMF1,SBMF2,SBMF3,Motrunich04}, renormalization group~\cite{RG} and numerical calculations~\cite{Watanabe2004,Chen2013,Ye2016,Kim2019,Mazumdar2016,Kivelson2020} have mainly focused on superconductivity and its pairing symmetry. Nevertheless, a systematic study of the correlated phases emerging from doped Mott insulators is a much needed endeavor, where many open questions are of equivalent importance as the intensively investigated  square lattice~\cite{PatrickLeeRMP,Dagotto1994,Zaanen2015}. Motivated by the above, here we study the  emergent correlated phases obtained on doping the distinct Mott insulators, which appear in the intermediate and strong coupling regimes,  in the triangular lattice Hubbard model. We focus on hole doping the $t>0$ model in Eqn. \ref{Model} below.

\begin{figure}[tbp]
\begin{center}
\includegraphics[width=0.48\textwidth]{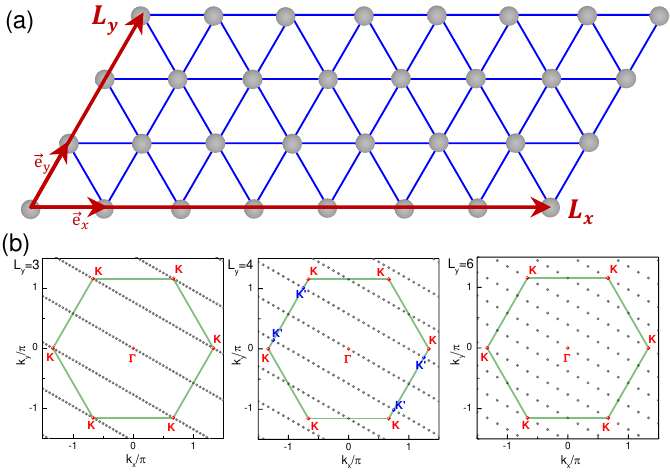}
\end{center}
\par
\renewcommand{\figurename}{Fig.}
\caption{(Color online) Triangular lattice and the corresponding Brillouin zone. (a) The triangular lattice is spanned by the primitive vectors $\mathbf{e_x}=(1,0)$ and $\mathbf{e_y}=(1/2, \sqrt{3}/2)$ with size $N=L_x\times L_y$. (b) The accessible momenta  (black dots) in the first Brillouin zone and the high-symmetry points are shown in (b) for $L_y=3$ (left), $L_y=4$ (middle) and  $L_y=6$ (right) cylinders. The number of independent momenta equals to size $N$, panel (b) shows three examples: $L_x=36$ (left), $L_x=30$ (middle) and $L_x=12$ (right).}
\label{Fig:Triangular}
\end{figure}

\begin{figure*}[tbp]
\begin{center}
\includegraphics[width=1\textwidth]{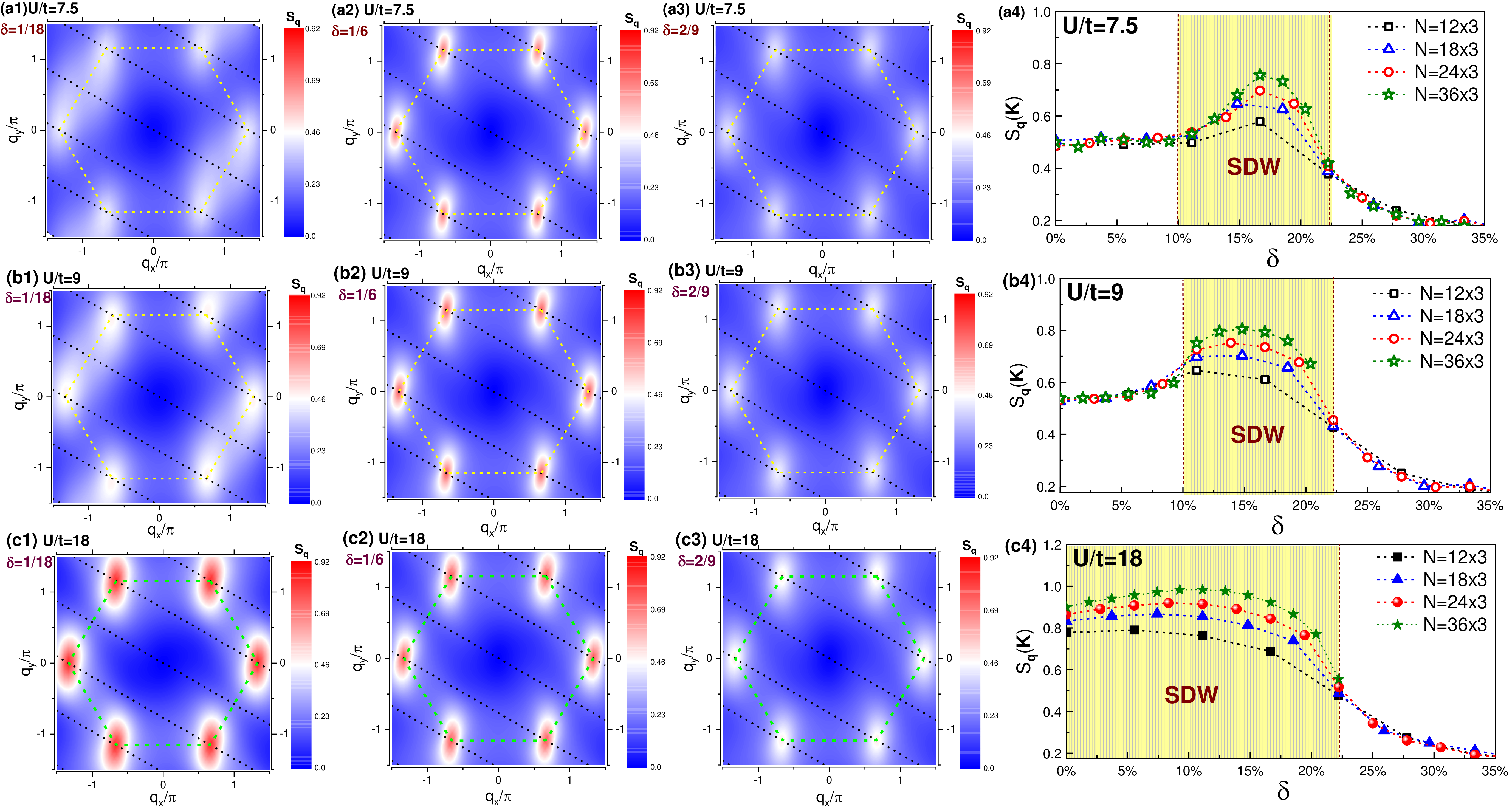}
\end{center}
\par
\renewcommand{\figurename}{Fig.}
\caption{(Color online) The static spin structure factor $S_\mathbf{q}(\mathbf{Q})$  as a function of hole doping  concentration $\delta$ on $L_y=3$ cylinders. Panels (a,b,c) show the contour plot of $S_\mathbf{q}(\mathbf{Q})$  for $U/t=7.5$ (a1-a4) , $U/t=9$ (b1-b4) and $U/t=18$ (c1-c4) with different $\delta$. From left to right in each row, we consider three typical hole doping concentrations: $\delta=1/18$ (a1,b1,c1), $\delta=1/6$ (a2,b2,c2) and $\delta=2/9$ (a3,b3,c3). The black dots represent the accessible momenta in the  Brillouin zone (dashed line) for $N=24\times3$ cylinders, and the contour plot is created by using triangulation interpolation.  Panels (a4) , (b4) and (c4) show $S_\mathbf{q}(\mathbf{Q})$ at $\mathbf{Q}=\mathbf{K}$ as a function of $\delta$ for $U/t=7.5$, $U/t=9$ and $U/t=18$, respectively.  The bond dimension of such calculation is set up to 6000$\sim$10,000. }
\label{Fig:SDW}
\end{figure*}

\begin{figure*}[tbp]
\begin{center}
\includegraphics[width=0.95\textwidth]{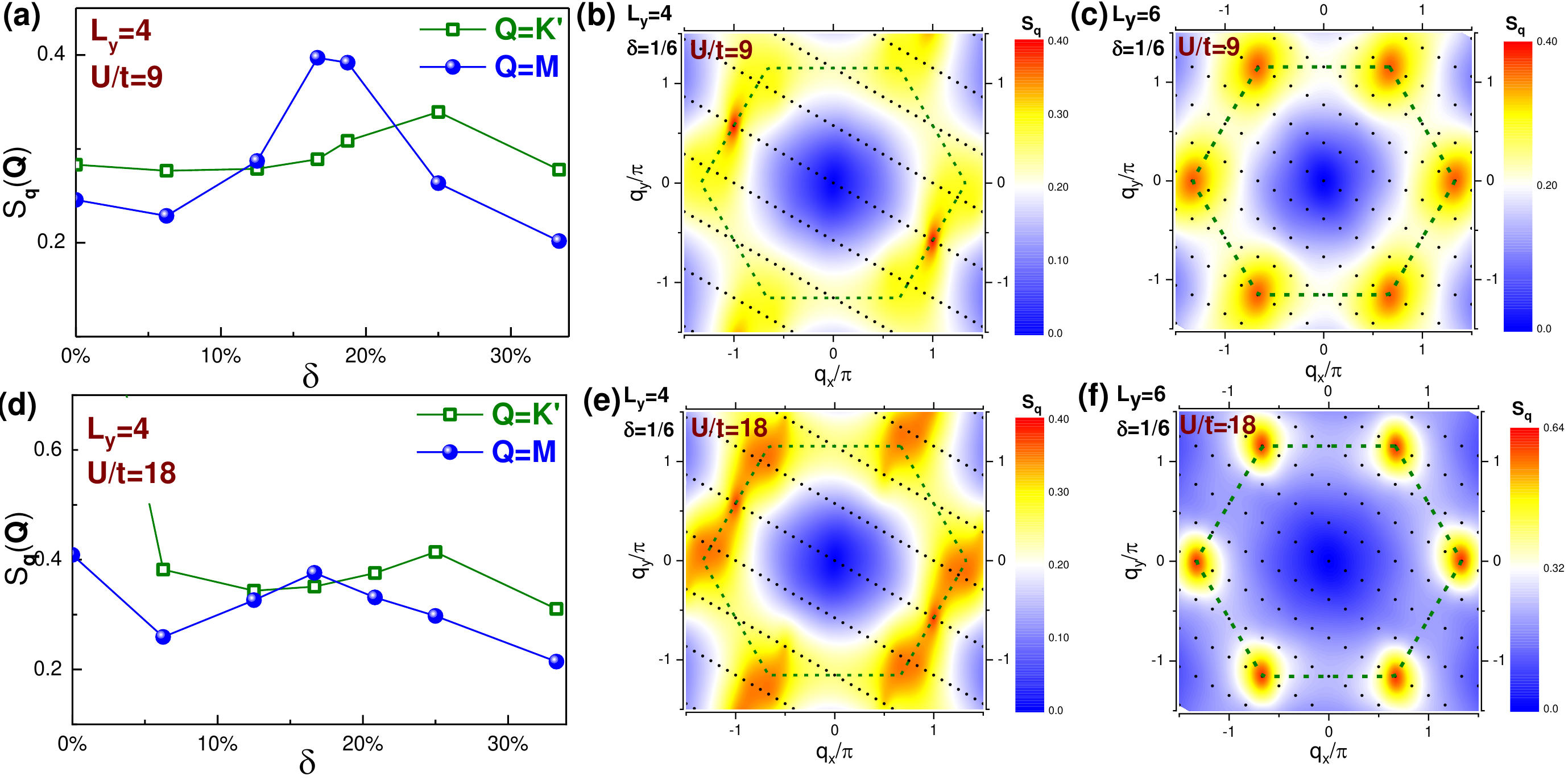}
\end{center}
\par
\renewcommand{\figurename}{Fig.}
\caption{(Color online) The static spin structure factor $S_\mathbf{q}(\mathbf{Q})$  on $L_y=4$ and $L_y=6$ cylinders.  Panel (a) and (d) show $S_\mathbf{q}$ as a function of hole doping $\delta$ for $U/t=9$ and $U/t=18$ at momentum $\mathbf{M}$ and  $\mathbf{K^\prime}$ on $N=24\times4$ cylinders. Panel (b) and (e) show the contour plot of $S_\mathbf{q}(\mathbf{Q})$  for  $U/t=9$ (b) and  $U/t=18$ (e) with $\delta=1/6$ and  $N=24\times4$ . Panel (c) and (f) show the similar plot of  $S_\mathbf{q}(\mathbf{Q})$ to (b) and (e) but on  $L_y=6$ cylinders with $L_x=12$. The black dots represent the accessible momenta points in the  Brillouin zone (dashed line) and the contour plot is created by using triangulation interpolation. The bond dimension of these calculation is set up to 30,000$\sim$35,000.
}
\label{Fig:4LSDW}
\end{figure*}

\section{Model and Method}
We consider the doped Hubbard model on a triangular lattice described by
\begin{equation}\label{Model}
H= - t\sum\limits_{\langle {i,j} \rangle, \sigma}   (c_{i\sigma}^\dag c_{j\sigma} + h.c. )+U\sum\limits_i {n_{i\uparrow}n_{i\downarrow}},
\end{equation}
where $\langle {i,j} \rangle$ denotes the nearest neighbor links, $c_{i\sigma}^{\dagger}$ ($c_{i\sigma}$ )  and $n_{i\sigma }$ represent the electron creation (annihilation) operators and number operators at site $i$ with spin $\sigma$ ($\sigma$ = $\uparrow,\downarrow$), respectively, and we take $t>0;\,U>0$. We perform a comparative study of the doped Mott insulators with distinct emergent spin backgrounds~\cite{Mishmash2015,Shirakawa2017,Szasz2018,Szasz2021,Chen2021,Wietek2021}: (i) a quantum disordered  spin background (``spin liquid'') that emerges at intermediate coupling strength and (ii) the magnetic ordered spin background with larger coupling strength. In this work, we mainly focus on the hole doped side and identify the nature of various doping-induced phases.   Considering the possible shift of the intermediate phase boundaries with system size~\cite{Mishmash2015,Shirakawa2017,Szasz2018,Szasz2021,Chen2021,Wietek2021}, we choose different typical parameters of $U/t$ for the doped magnetic disordered spin background and the doped magnetic ordered spin background.
 
Due to the lack of well controlled theoretical methods in two-dimensional strongly correlated systems, quasi-one-dimensional systems have become a good starting point allowing two-dimensional characteristics to emerge on growing the degrees of freedom compared to one dimension. More crucially, they allow to perform accurate numerical density matrix renormalization group (DMRG) simulations, which have proved to be one of the most powerful numerical methods for strongly correlated systems such as the doped Mott insulators on the square lattice.
Since the computational cost of DMRG~\cite{DMRG1,DMRG2} increases exponentially with system width, we focus on cylinders similar to earlier DMRG studies on the square lattice.  The cylinder is spanned by  vectors $L_x \mathbf{e_x}=L_x (1,0)$ and $L_y\mathbf{e_y}=L_y (1/2, \sqrt{3}/2)$ with circumference $L_y$, as illustrated in Fig.~\ref{Fig:Triangular} (a).

Figure~\ref{Fig:Triangular} (b) depicts the corresponding Brillouin zone for $L_y=3,4,6$ cylinders, and the black dots represent the accessible momenta points on systems with size $N=L_x\times L_y$. Due to the fact that the spin long-range ordered phase  becomes gapped for even $L_y$~\cite{Affleck1989}, it requires odd $L_y$ and integral multiple of 3 for both $L_x$ and $L_y$ in order to capture the nature of the $120^{\circ}$ N\'eel order.  In particular, $\mathbf{K}$ points are inaccessible on cylinders with $L_y=4$,  as shown in Fig.~\ref{Fig:Triangular} (b),  we can only access $\mathbf{K^\prime}$ points, which are the closest momentum to  $\mathbf{K}$. In the present work,  we mainly focus on $L_y=3$ and $L_y=4$ but also compare with $L_y=6$ cylinders when identifying the wave vectors of the spin density waves (SDWs).

Depending on system size and physical quantity, the bond dimension is set up to D=45, 000 when implementing  $U(1)\times U(1)$ symmetry in the DMRG program, and up to D=23,000 spin multiplets when implementing  $U(1)\times SU(2)$ symmetry ($\sim$ D=69,000 in $U(1)\times U(1)$ program).  The cylinder length is pushed up to $L_x=72$, and we also need to point out that the longer length also requires much larger bond dimension to get converged results for quantities such as the chiral correlations.  For example, for the chiral-chiral correlations on $L_y=4$ cylinders, we find that, at light doping, the converged measurement can only be obtained at D=23, 000 with using $U(1)\times SU(2)$ symmetry, which roughly corresponds to D$\approx$69,000 when using $U(1)\times U(1)$ symmetry in the program.  In particular, for the fast decay of the correlation functions, both power-law and exponential function could fit the data well; to see it more clearly, we present both semi-logarithmic and double-logarithmic plots for the same data to compare.

\section{Results}
 
\subsection{Evolution of Spin Correlations with Hole Doping}

We begin with probing the ground state properties of the model Hamiltonian (\ref{Model}) as a function of hole doping $\delta$ in the spin channel by examining the spin structure factor
\begin{equation}
S_\mathbf{q}(\mathbf{Q}) =\frac{1}{N}\sum_{i,j} {\left\langle {S_i^zS_j^z} \right\rangle e^{i \mathbf{Q}\cdot(\mathbf{r_i}-\mathbf{r_j})}}.
\end{equation}
Here, we have confirmed that $\left\langle {S_i^z} \right\rangle$ is vanishingly small on each site. In our calculation, the hole doping is realized by removing equal number of spin-up and spin-down electrons, and we target the sector with total spin $S^z=0$. Figures~\ref{Fig:SDW} (a1-a3), (b1-b3), (c1-c3) show the contour plot of $S_\mathbf{q}$ at three typical hole doping concentrations for $U/t=7.5$, $U/t=9$ and $U/t=18$, respectively. The spin structure factor for $U/t=7.5$ and $U/t=9$ exhibit similar behavior.

{\em Spin correlations at intermediate coupling  for $L_y=3$:} For  $U/t=7.5$ and $U/t=9$, $S_\mathbf{q}$ is featureless at light doping [see Figs.~\ref{Fig:SDW} (a1,b1) for $\delta=1/18$], consistent with a spin disordered phase.
With further doping, $S_\mathbf{q}$ exhibits sharp peaks at $\mathbf{Q=K}$ for moderate doping [see Figs.~\ref{Fig:SDW} (a2 ,b2) for $\delta=1/6$], suggesting a doping induced commensurate SDW. The intermediate SDWs are finally suppressed on further increasing doping beyond $20\%$ [see Figs.~\ref{Fig:SDW} (a3,b3) for $\delta=2/9$]. To probe the evolution of SDWs with hole doping, we keep track of $S_\mathbf{q}(\mathbf{K})$ as a function of $\delta$, as shown in Figs.~\ref{Fig:SDW} (a4, b4), $S_\mathbf{q}(\mathbf{K})$ is independent of  {$L_x$} for $\delta\lesssim10\%$ at intermediate coupling, consistent with a nonmagnetic spin background. However, on further increasing doping to $\delta\approx10\%\sim20\%$,   $S_\mathbf{q}(\mathbf{K})$ is significantly enhanced and its height also increases with system size, indicating the doping induced SDW, which finally disappears at $\delta \gtrsim 20\%$.

{\em Spin correlations at strong coupling  for $L_y=3$:} In contrast, on doping holes into the strong coupling $U/t=18$ model, a 120$^{\circ}$ N\'eel ordered spin background with sharp peaks in $S_\mathbf{q}(\mathbf{K})$ survives until $\delta \approx 20\%$ for $L_y=3$ [see Figs.~\ref{Fig:SDW} (c1-c3)]. Furthermore,  the height of these peaks  also increase with system sizes [see Fig.~\ref{Fig:SDW} (c4)], indicating  that the commensurate SDW order remains robust against hole doping in the strong coupling limit.
At $\delta > 20\%$, the spin correlations become short ranged and are indistinguishable for all coupling strengths.

{\em Spin correlations for $L_y=4$:} For wider cylinders with $L_y=4$, we find the spin backgrounds at $\delta\lesssim10\%$ also resemble the ones at half filling, as shown in Fig.~\ref{Fig:4LSDW}(a) and (d). Although the momenta $\mathbf{K}$ are inaccessible for $L_y=4$ and the spin ordered phase becomes spin gapped due to the even-leg effect,  the intensity of  $S_\mathbf{q}$ is concentrated at  the momentum closest to $\mathbf{K}$ (i.e., $\mathbf{K^\prime}$) in the strong coupling limit (at $\delta\lesssim10\%$). These facts indicate the nature of spin background at light doping is mainly determined by coupling strength $U/t$.
At moderate doping $\delta\thickapprox 10\%\sim20\%$, the commensurate SDWs exhibits competing wave vector $\mathbf{M}$, as shown in  Figs.~\ref{Fig:4LSDW} (b,e). To show it more clearly, we study $S_\mathbf{q}$ at $\mathbf{M}$ and in the vicinity of $\mathbf{K}$ as a function of hole doping, as shown in Figs.~\ref{Fig:4LSDW} (a,d), the intensity of $S_\mathbf{q}(\mathbf{M})$ is enhanced at moderate doping, while the SDWs are finally suppressed for larger doping $\delta\gtrsim20\%$.

We also notice the competing wave vectors of the SDW, which is determined by the coupling strength $U/t$. The SDW with wave vector $\mathbf{Q=M}$ is dominant at moderate coupling strength, however, with the increase of coupling strength $U/t$, the SDW with $\mathbf{Q=K^\prime}$ becomes competitive, as indicated from Figs.~\ref{Fig:4LSDW} (a) and (d). We also have checked the $t-J$ model, which corresponds to the effective Hamiltonian of Hubbard model in the $U/t\rightarrow\infty$ limit, and confirmed that the dominant wave vector of SDW is $\mathbf{Q=K^\prime}$ in the strong coupling limit. Such competition might be induced by the special feature of $L_y=4$ cylinders, where  $\mathbf{K}$ points are inaccessible. To further confirm it, we also check $L_y=6$ cylinders and find  that when both the momenta $\mathbf{K}$ and $\mathbf{M}$ are accessible, the commensurate SDW with wave vector $\mathbf{Q=K}$ is dominant at these dopings, as shown in Fig.~\ref{Fig:4LSDW} (c) and (f).

\begin{figure}[tbp]
\begin{center}
\includegraphics[width=0.42\textwidth]{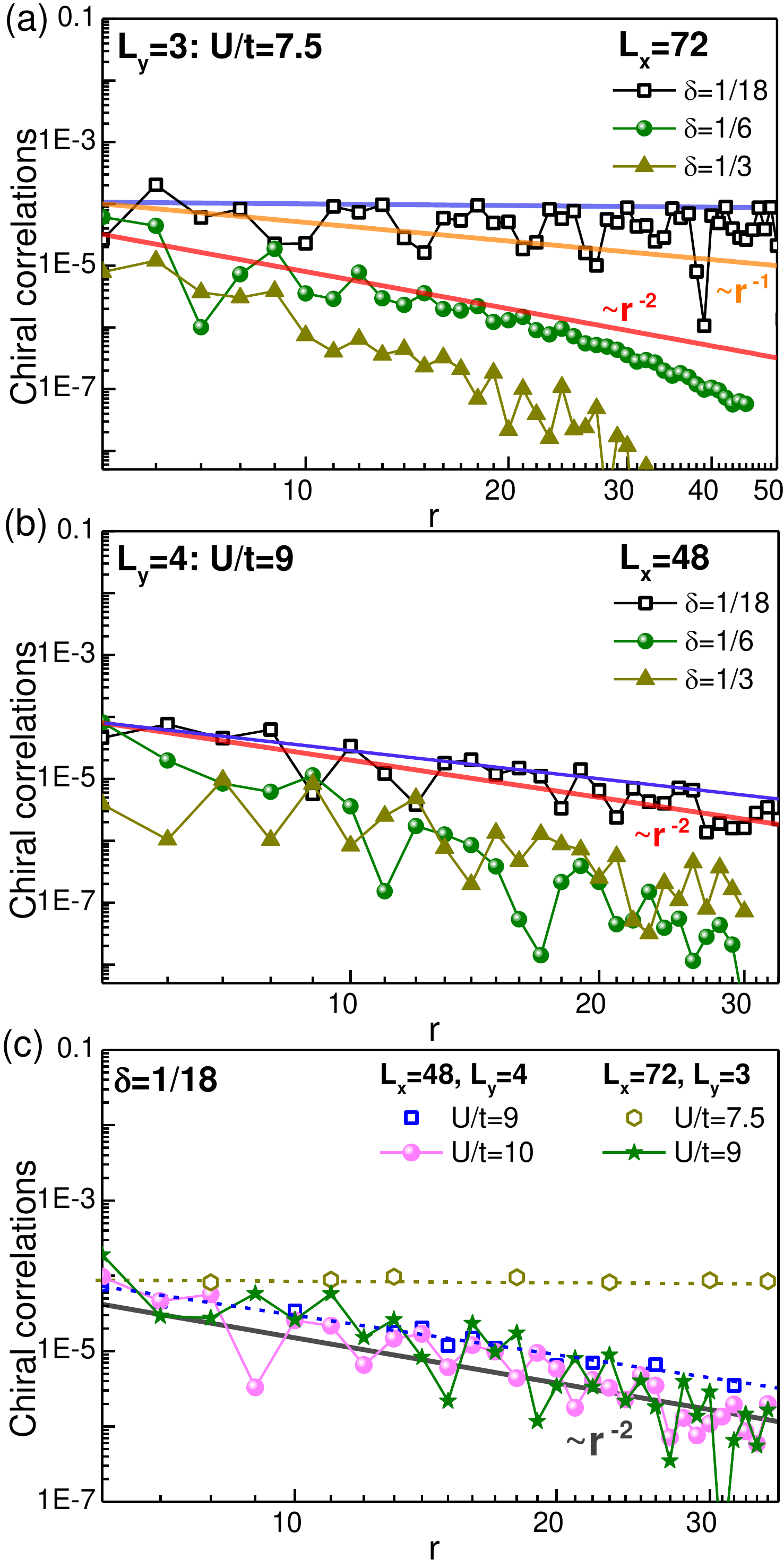}
\end{center}
\par
\renewcommand{\figurename}{Fig.}
\caption{(Color online) Chirality correlators. The chiral-chiral correlations  as a function of distance $r$ for three typical hole doping at $U/t=7.5$ on $N=72\times3$ cylinders (a) and at $U/t=9$ on $N=48\times4$ cylinders (b). The solid lines represent the power-law function of $r^{-2}$ (red) and $r^{-1}$ (orange). The blue solid lines correspond to the decay of chiral correlations.  The bond dimension of such calculation  is set up to 36,000$\sim$60,000 for $L_y=3$ in (a) and up to 36,000$\sim$69,000 for $L_y=4$ in (b). Panel (c) shows additional results at $U/t=9$ on $N=72\times3$ cylinders and at $U/t=10$ on $N=48\times4$ cylinders, we also include the results (a-b) to compare.}
\label{Fig:Chiral}
\end{figure}

\subsection{Evolution of Spin Chirality with Hole Doping}

To further investigate the nature of magnetic disordered spin background at $\delta\lesssim10\%$ with intermediate coupling strength, we examine the spin chiral order by computing the chiral-chiral correlations $|\langle C_{\bigtriangleup_i} C_{\bigtriangleup_{i+r}}\rangle|$,
where the chiral operator $C_{\bigtriangleup_i}={{\mathbf{S}}_{{i_1}}} \cdot \left( {{{\mathbf{S}}_{{i_2}}} \times {{\mathbf{S}}_{{i_3}}}} \right)$ is defined on the triangle formed by three nearest neighboring sites $i_1$, $i_2$ and $i_3$. As shown in Fig.~\ref{Fig:Chiral} (a) for $L_y=3$ cylinders and Fig.~\ref{Fig:Chiral} (b) for $L_y=4$ cylinders at three typical hole doping concentrations: $\delta=1/18$, $\delta=1/6$ and $\delta=1/3$.

At light doping $\delta\lesssim10\%$, the chiral correlations decay much slower than $r^{-1}$ [see the orange line in Fig.~\ref{Fig:Chiral} (a)] for $L_y=3$ at $U/t=7.5$ and slightly slower than $r^{-2}$ [see the red line in Fig.~\ref{Fig:Chiral} (b)] for $L_y=4$ at $U/t=9$.  The blue lines in Figs.~\ref{Fig:Chiral} (a-b) follow the amplitude of the chiral correlations, which can directly compare with the decay rate $r^{-1}$ (orange line) and $r^{-2}$ (red line), respectively. These results suggest the (quasi-)long range chiral order  at light doping. We also check $U/t=9$ for $L_y=3$ and $U/t=10$ for $L_y=4$, as shown in Fig.~\ref{Fig:Chiral} (c), which exhibit similar quasi-long-range chiral correlations decaying in a power-law fashion. Here we point out that $U/t=9$ and $U/t=10$ are similar for $L_y=4$, while the chiral correlations at $U/t=9$ decay faster than $U/t=7.5$ but comparable to $r^{-2}$, which indicates the increase of the coupling strength $U/t$ would suppress the chiral correlations, but the spin background is still nonmagnetic according to the featureless spin structure factor $S_\mathbf{q}$ in Fig.~\ref{Fig:SDW} (a1). When we increase the doping to $\delta \gtrsim 10\%$, the chiral correlations are strongly suppressed and decay faster than $r^{-2}$, as shown in in Figs.~\ref{Fig:Chiral} (a-b).

We also point out that the chiral correlations have sign fluctuations  for $L_y=3$ but not for $L_y=4$ for half filling, consistent with Refs.~\cite{Szasz2018,Szasz2021}, however, at light doping, both cases exhibit the change of sign with distance $r$. In particular, we notice that, the claim of absence of chiral spin liquid at half-filling for $L_y=3$ in Ref.~\cite{Peng2021} also conflicts with other DMRG studies~\cite{Szasz2018,Szasz2021,Knap2022}.

\begin{figure}[tbp]
\begin{center}
\includegraphics[width=0.45\textwidth]{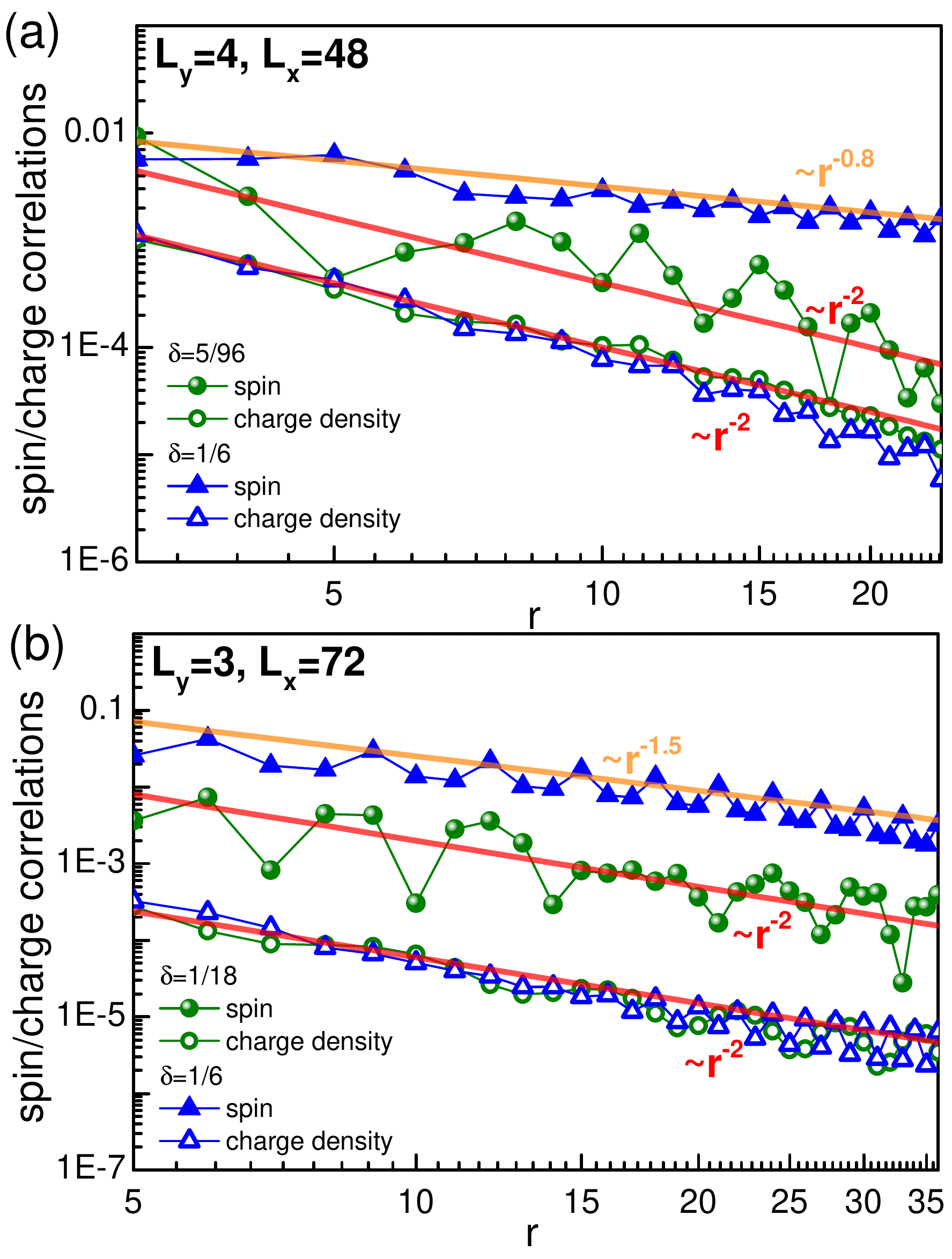}
\end{center}
\par
\renewcommand{\figurename}{Fig.}
\caption{(Color online) Spin-spin correlations and charge density-density correlations.  Panels (a) and (b) show the comparison between the spin-spin correlations and charge density-density correlations for $U/t=9$ on $L_y=4, L_x=48$ and $L_y=3, L_x=72$ cylinders. We consider two typical examples at light doping and moderate doping.  Both correlations decay in a comparable rate at light doping, however, at moderate doping,  the spin-spin correlations decay much slower than the charge density-density correlations and also with much larger amplitude. The bond dimension of such calculation  is set up to 60,000 for light doping and up to 36,000 for moderate doping $\delta\lesssim10\%$.}
\label{Fig:SDWCDW}
\end{figure}

\subsection{SDW vs. CDW}\label{SDWvsCDW}
For the doped Mott insulators on square lattice, the doped charge would suppress the N\'eel order at light doping, and the CDW, such as the unidirectional stripes, would emerge  around $\delta \approx 1/8$. Here, on the triangular lattice, from the above results of the spin structure factor $S_\mathbf{q}(\mathbf{Q})$ and the chiral-chiral correlations $|\langle C_{\bigtriangleup_i} C_{\bigtriangleup_j}\rangle|$, we find the nature of the spin background is robust against hole doping and also find the robust spin density waves at moderate doping  $\delta\approx10\%\sim20\%$. In this section, we directly compare the spin-spin correlations $|\left\langle {S_i^zS_{i+r}^z} \right\rangle|$ with the charge density-density correlations $|\left\langle {n_i n_{i+r}} \right\rangle- \langle n_i\rangle \langle n_{i+r}\rangle|$ to examine the dominant correlations.

At light doping  $\delta\lesssim10\%$, as shown in Figs.~\ref{Fig:SDWCDW} (a-b), the spin-spin correlations decay slightly slower than the charge density-density correlations or with comparable rate, the amplitude of the spin-spin correlations is also larger. The red lines in Figs.~\ref{Fig:SDWCDW} represent the decay rate $\sim r^{-2}$, which can be used for guidance. With increasing the hole doping to  moderate level $\delta\approx10\%\sim20\%$. The charge density-density correlations are almost unchanged or slightly suppressed [see  Figs.~\ref{Fig:SDWCDW} (a-b)], however, the spin-spin correlations are significantly enhanced with both much slower decay rate and larger amplitude. The orange lines in  Figs.~\ref{Fig:SDWCDW} suggest the decay rate of the spin-spin correlations close to $\sim r^{-1.5}$ for $L_y=3$ and is further enhanced to $\sim r^{-0.8}$ for wider cylinders $L_y=4$.  Since the power-law decay $\sim r^{-\alpha}$ with $\alpha<2$ suggests the diverged susceptibility towards  2D, these observations indicate that the SDW would be dominant over the CDW for the doped Mott insulators on the triangular lattice.

Here we point out that the recent work \cite{Peng2021} also shows similar findings, i.e., the comparable decay rate of both correlations and the larger amplitude of the spin correlations. Although both correlators can be fitted by the power-law function with close exponents similar to our findings in Figs.~\ref{Fig:SDWCDW} (a-b), their interpretations that the spin correlations are  exponentially decaying while the charge density correlations are  power-law in Ref.~\cite{Peng2021} are inconsistent with our results.

\begin{figure*}[tbp]
\begin{center}
\includegraphics[width=0.95\textwidth]{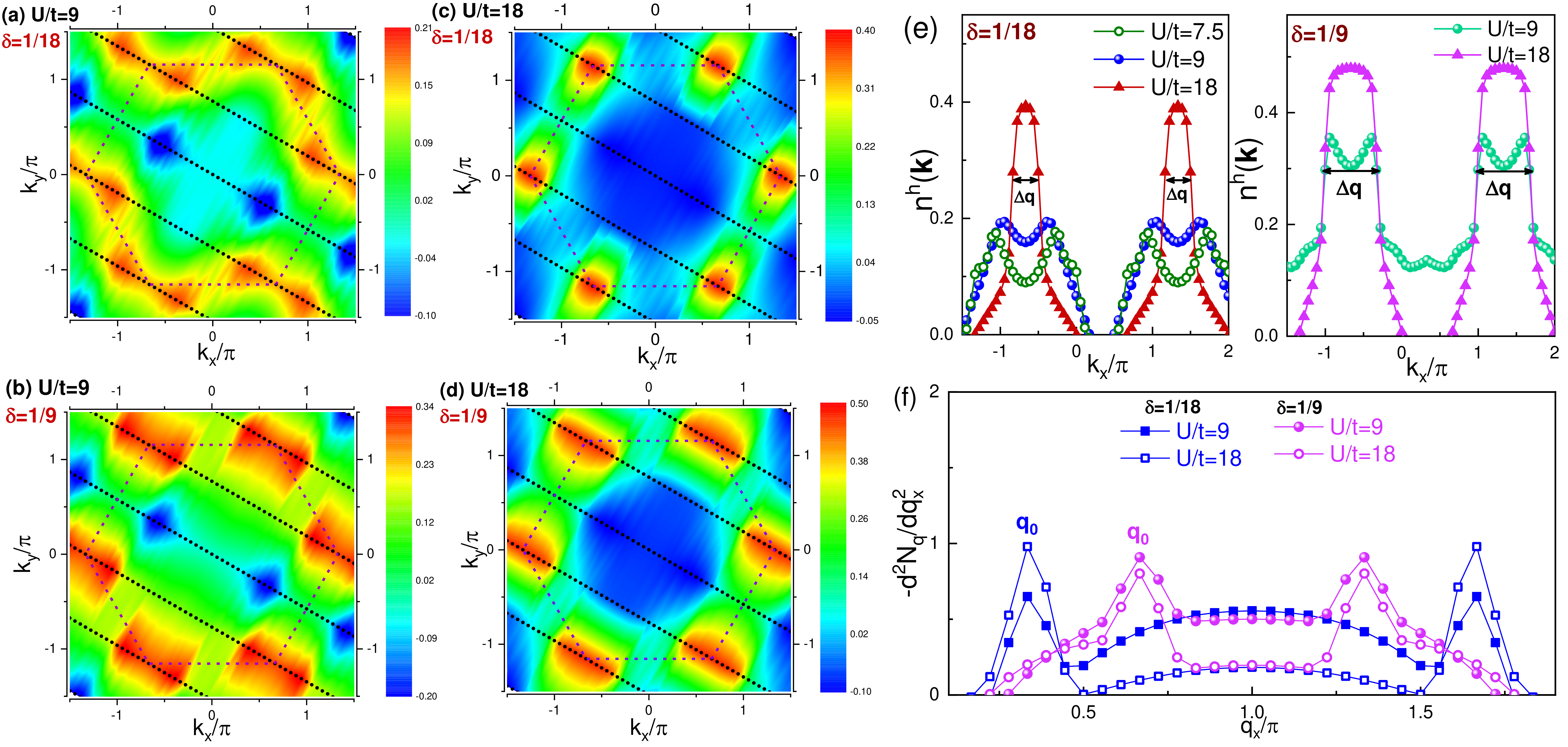}
\end{center}
\par
\renewcommand{\figurename}{Fig.}
\caption{(Color online) The properties of doped holes - Fermi surface and charge structure factor. (a-d) The contour plot of electron momentum distribution $n^h(\mathbf{k})$ for  $U/t=9$ (a-b) and $U/t=18$ (c-d) at doping $\delta=1/18$ and $\delta=1/9$. The black dots represent the accessible momenta points in the  Brillouin zone, and the contour plot is created by using triangulation interpolation. Panel (e) show the cuts of $n^h(\mathbf{k})$ through $\mathbf{K}$ at doping $\delta=1/18$ (left) and $\delta=1/9$ (right), respectively. Panel (f) shows the second order derivatives of the density structure factor $N_q$, the peaks at $q_0$ indicate the wave vector of charge modulations. Here we consider $N=36\times3$ cylinders.  The bond dimension of such calculation is set up to 10,000$\sim$12,000.
}
\label{Fig:Charge2}
\end{figure*}

\begin{figure*}[tbp]
\begin{center}
\includegraphics[width=0.95\textwidth]{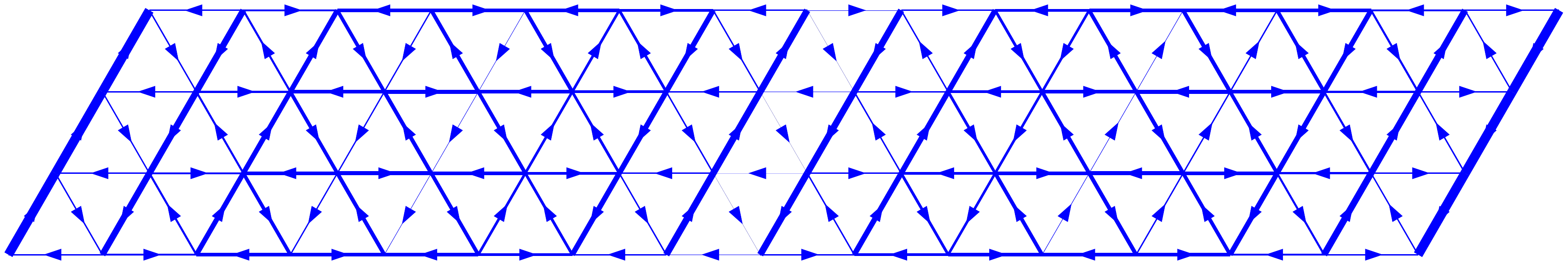}
\end{center}
\par
\renewcommand{\figurename}{Fig.}
\caption{(Color online) The charge current pattern in the chiral metal phase. This pannel shows the pattern of $I^{c}_{\langle ij\rangle}$ at $\delta=1/20$ for $U/t=9$ on $L_y=4$ cylinders, the width of each bond is proportional to the current magnitude while the arrow indicates the current direction. Here we consider $N=16\times4$ cylinders. The bond dimension of such calculation is set up to $12,000$.}
\label{Fig:ChargeCurrent}
\end{figure*}

\begin{figure}[tbp]
\begin{center}
\includegraphics[width=0.48\textwidth]{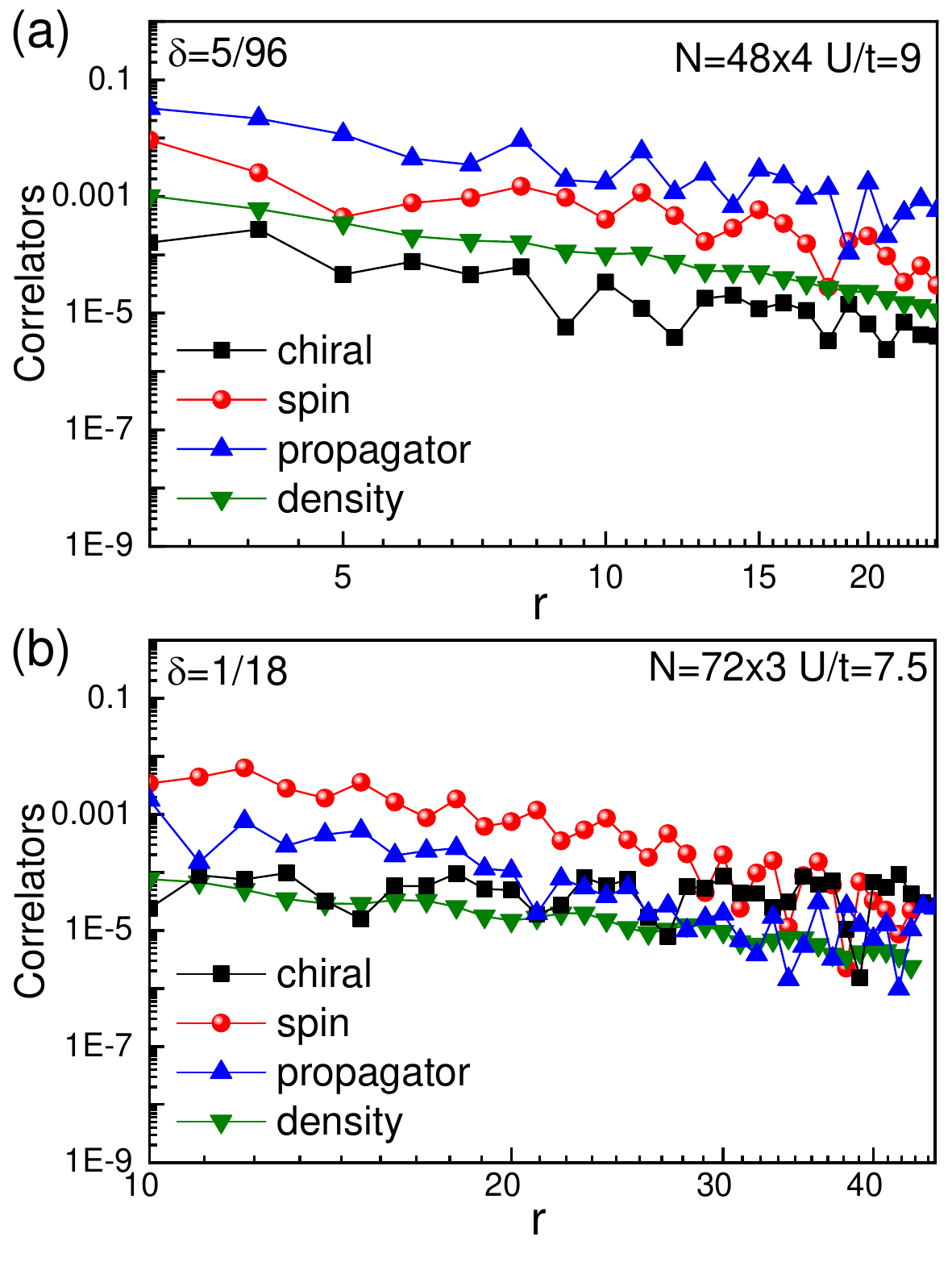}
\end{center}
\par
\renewcommand{\figurename}{Fig.}
\caption{(Color online) Real-space correlators at light doping: chiral-chiral correlations, spin-spin correlations, charge density-density correlations, single particle propagators.  Panel (a) and (b) show different correlators for $U/t=9$ on $L_y=4$ and for $U/t=7.5$ on $L_y=3$ cylinders, respectively. Here we focus on the doping.}
\label{Fig_CorrelatorsCM}
\end{figure}

\begin{figure*}[tbp]
\begin{center}
\includegraphics[width=0.95\textwidth]{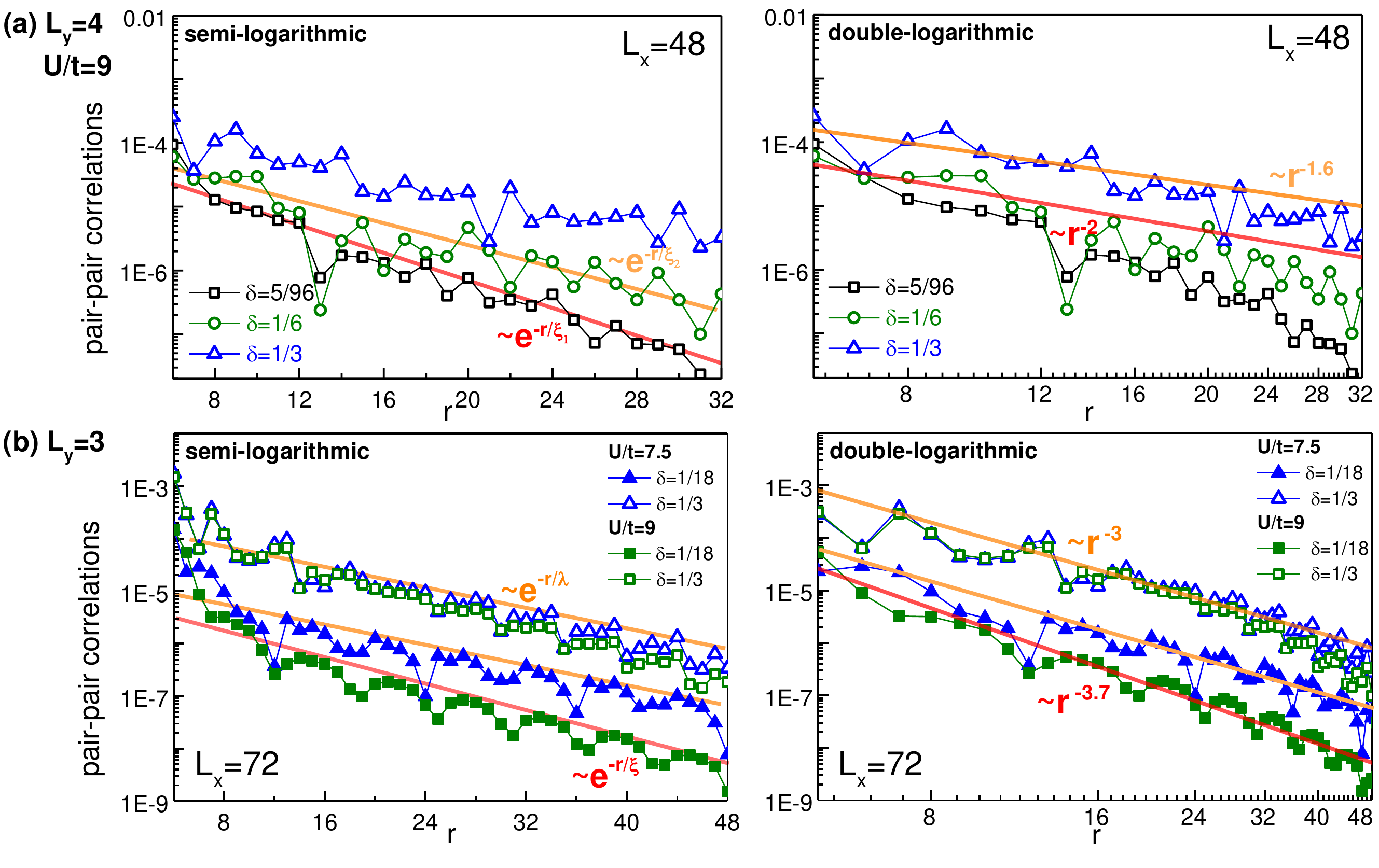}
\end{center}
\par
\renewcommand{\figurename}{Fig.}
\caption{(Color online) The pair-pair correlations. Panels (a-b) show the pair-pair correlations in singlet channels for $U/t=9$  on  $L_y=4, L_x=48$ (a) and  $L_y=3, L_x=72$ (b) cylinders. The left panels in each figure show the plots in semi-logarithmic scale, while the right panels are the plots in double-logarithmic scale. The bond dimension of such calculation is set up to 60,000.
}
\label{Fig:SC}
\end{figure*}

\subsection{Hole Pockets Evolution with  Doping}
In the charge channel, we examine the properties of doped holes by measuring {the hole momentum distribution function $n^h(\mathbf{k})$,  which can be extracted from the change of electron momentum distribution after doping:
\begin{equation}
n^h(\mathbf{k})\equiv n_{0} (\mathbf{k})-n_\delta(\mathbf{k}),
\end{equation}
 where
 \begin{equation}
 n_\delta(\mathbf{k})=\frac{1}{N}\sum_{i,j,\sigma }\langle c_{i\sigma }^\dag c_{j\sigma }\rangle e^{i\mathbf{k}\cdot(\mathbf{r_i}-\mathbf{r_j})}
  \end{equation}
  represents the electron momentum distribution at hole doping $\delta$, $n_0(\mathbf{k})$ corresponds to half filling  at the same coupling strength $U/t$. }

At light hole doping $\delta<10\%$,  we find the following contrasting outcomes depending on whether one is at strong or intermediate coupling. At strong coupling, the doped holes  form small pockets around momenta $\mathbf{K}$ [see Fig.~\ref{Fig:Charge2} (c) for $U/t=18$],  while at intermediate coupling  each hole pocket splits into two parts  [see Fig.~\ref{Fig:Charge2} (a) for $U/t=9$].  To see this more clearly, we show the cuts of $n^h(\mathbf{k})$ across the hole pockets in Fig.~\ref{Fig:Charge2} (e). Strikingly, at strong coupling there is a sharp drop in $n^h(\mathbf{k})$ characterizing a well defined Fermi momentum of holes in the spin ordered background, while at intermediate coupling there is a broad momentum distribution for lightly doped holes. The latter is indicative of  fractionalized spin and charge excitations - although the total momentum of the spin and charge is conserved, the momentum shared between separated charge and spin excitations would lead to the absence of a well defined Fermi momentum for holes~\cite{ZZ2018a,ZZ2018b,ZZ2015}. Here we also point out that the splitting of hole pockets at  intermediate coupling is independent of $L_x$, and gradually disappears with increasing coupling strength to the strong coupling regime, as shown in the left panel in Fig.~\ref{Fig:Charge2} (e).

\subsection{Chiral metallic phase}

We also compared $n^h(\mathbf{k})$ with charge density structure factor $N_\mathbf{q}$ to further confirm it.  $N_\mathbf{q}$ is defined by the Fourier transformation of density-density correlations, i.e.,
\begin{equation}
N_\mathbf{q} =\frac{1}{N}\sum_{i,j} {(\left\langle {n_i n_j} \right\rangle- \langle n_i\rangle \langle n_j\rangle )e^{i \mathbf{q}\cdot(\mathbf{r_i}-\mathbf{r_j})}}.
\end{equation}
As shown in Fig.~\ref{Fig:Charge2} (f), the peaks in the second order derivative of $N_\mathbf{q}$ characterize wave vectors of the charge modulations $\mathbf{q_0}$,
 or equivalently, the scattering between holes near the Fermi surface with momentum difference $\Delta \mathbf{q}$.
For a Fermi liquid state that appears to be present in the strong coupling regime, $\mathbf{q_0}=\Delta \mathbf{q}$, where the corresponding charge modulations (see Appendix~\ref{AppedixCM}) can be attributed to the intra-pocket scattering.

A striking observation is that in the anomalous chiral metal (the lightly doped intermediate coupling state) we  find  charge modulations with the {\em same} wave vectors $\mathbf{q_0}$ as in the strong coupling limit. However, unlike in the strong coupling limit, a well defined Fermi momentum for doped holes is lacking [see Fig.~\ref{Fig:Charge2} (e)]. The significant difference between $N_\mathbf{q}$ and $n^h(\mathbf{k})$ suggests the spin and charge are no longer confined together in the anomalous chiral metal, consistent with doping a spin liquid. At larger doping $\delta \gtrsim 10\%$, a well defined hole pocket is reconstructed even at intermediate coupling strength around  $\mathbf{K}$ [see Fig.~\ref{Fig:Charge2} (b)], while for the strong coupling strength, the original hole pockets gradually increase with doping [see Fig.~\ref{Fig:Charge2} (d)]. Both cases have the same Fermi momentum, as illustrated by the jump in $n^h(\mathbf{k})$ in Fig.~\ref{Fig:Charge2} (e). The momentum difference $\Delta \mathbf{q}$ between holes near Fermi surface exactly matches the peak $\mathbf{q_0}$ in the  second order derivative of the charge structure factor, $N_\mathbf{q}$, i.e., $\mathbf{q_0}=\Delta \mathbf{q}$.

Moreover, we compare various real-space correlators to further examine the chiral metallic phase. As shown in Fig.~\ref{Fig_CorrelatorsCM} (a) for $L_y=4$ cylinders, we find the single particle propagator decays in a power-law fashion with a slower decay rate compared with other correlations,
suggesting the robust nature of chiral metallic phase on $L_y=4$ cylinders. However, we also point out that we find complex feature on  $L_y=3$ cylinders [see Fig.~\ref{Fig_CorrelatorsCM} (b)]: (i)the spin chirality order is dominant over other correlations for long distance on  $L_y=3$ cylinders, indicating a robust time-reversal symmetry breaking phase; (ii) the single particle propagator decays relatively faster than $L_y=4$ but tends to saturate at longer distance and decays comparable with chiral correlations; (iii)the spin-spin correlations ($\sim r^{-2.8}$) decay in a comparable rate with the pair-pair correlations [see the next section and Fig.~\ref{Fig:SC} (b)] and exhibiting spatial oscillations, which suggests  the pair-density-wave (PDW) pattern is locally robust but the quasi-long-range order is insignificant. Therefore the $L_y=3$ cylinder exhibits the competition between the chiral metal and local PDW.

We further probe the nature of this chiral metal state by examining the bond charge current pattern, which is defined by $I^{c}_{\langle ij\rangle}\equiv -\text{i}\sum_{\sigma}\langle{c}_{i\sigma}^{\dag}{c}_{j\sigma}-h.c.\rangle$. A typical example of the current pattern at $\delta=1/20$ for $U/t=9$ is shown in Fig.~\ref{Fig:ChargeCurrent}, where the width of each bond is proportional to the current magnitude and the arrow indicates the current direction, the bond current pattern exhibits the translational symmetry breaking in the bulk along the horizontal direction, though we also notice the existence of domains in the pattern. We find $L_y=3$ cylinders exhibit the same feature with $L_y=4$, and leave such results in the Appendix~\ref{AppedixCC}.  Meanwhile, the time-reversal invariant spin current $I^{s}_{ij}\equiv -\text{i}\langle{S}_{i}^{+}{S}_{j}^{-}-S_{i}^{-}{S}_{j}^{+}\rangle$ is vanishingly small in the same phase, consistent with the existence of the spin chirality order for chiral metal. Here, we would like to mention that the chiral metal phase identified here is consistent with recent mean-field analysis of the doped Kalmeyer-Laughlin type chiral spin liquid\cite{KLstate}, where a chiral metal with unit cell doubling and staggered loop current order is proposed\cite{Song2020}. Ref.~\cite{Song2020} also points out the competition between the chiral metal and superconductivity, which is also observed here for $L_y=3$. But with the increase of cylinder width, we find the chiral metallic phase becomes dominant.

\subsection{Evolution of Superconducting Pair-Pair Correlations with Hole Doping}\label{SCpair}
Below we examine the superconductivity on hole doping by measuring the pair-pair correlations $D(r)\equiv \left\langle (\hat{\Delta}^{s,t}_{i})^{\dagger} \hat{\Delta}^{s,t}_{i+r}\right\rangle $, in which the Cooper pair operators in the singlet and triplet channels are defined by $\hat{\Delta}^s_{i} \equiv \frac 1 {\sqrt{2}}\sum_{\sigma}\sigma {c}_ {i_1, \sigma} { c}_ {i_2, -\sigma}$,  and $\hat{\Delta}^t_{i}  \equiv \frac 1 {\sqrt{2}}\sum_{\sigma} {c}_ {i_1,  \sigma} {c}_ {i_2,  -\sigma}$, respectively.  Here, we focus on the local pairing between the nearest sites $(i_1,i_2)$. We have fixed one bond in the pair-pair correlations at $i$ along $\mathbf{e_y}$ and measure its correlations with pairs along $\mathbf{e_y}$,$\mathbf{e_x}$ and $\mathbf{e_x-e_y}$, respectively, then we average the absolute value of pair correlations for a fixed distance.
Figures~\ref{Fig:SC} (a-b) show the pair-pair correlations for $U/t=7.5, 9$ for $L_y=3$ and $U/t=9$ for $L_y=4$ at typical doping levels on $L_y=3, 4$ cylinders, and $U/t=18$ gives similar results. For both intermediate coupling and strong coupling  models, we find the pairing strength in singlet channel are stronger than triplet channel, and thus we will focus on singlet pairing. We also notice that, at larger doping $\delta  > 20\%$, while the singlet
pairing has stronger correlations at longer distance, the triplet pairing becomes competitive with singlet pairing when further increasing $\delta$, particularly for wider systems.

In our quasi-one-dimensional setup, true long-range order in the pair correlation function $D(r)$ is forbidden by the Mermin-Wagner theorem. We therefore content ourselves with looking for slow power law decay ($D(r) \sim 1/r^\eta$ where $\eta \sim 1$  ), which, as we show, appears at the largest cylinder circumferences and at high doping. To gain the indication of superconductivity for 2D,  we look for power law decay with $\eta < 2$, which would lead to the divergence of superconductivity susceptibility. For the fast decay of the correlation functions, both power-law and exponential function could fit the data well, to see it more clearly, we present both semi-logarithmic and double-logarithmic plot for the same data to compare, as shown in Fig.~\ref{Fig:SC}.

For $L_y=4$ cylinders, as shown in Fig.~\ref{Fig:SC} (a), the pair-pair correlations $D(r)$ decay exponentially at light doping, while its amplitude and  decay length increase  with the increase of hole doping. At moderate hole doping $\delta\approx10\%\sim 20\%$,  $D(r)$ could be fitted by an exponential function with long decay length or a power-law function with relatively large exponent $\eta \gtrsim 2$, as shown in the semi-logarithmic and double-logarithmic plot in Fig.~\ref{Fig:SC} (a). Due to the large exponent, the power-law fitting and the exponential fitting would be very close, implying that the superconductivity is not dominant when the SDWs exist.  At $\delta\gtrsim 20\%$, the pair-pair correlations are further enhanced when the SDWs are suppressed. The double-logarithmic in the right panel of Fig.~\ref{Fig:SC} (a) shows $\eta \approx 1.6$ at $\delta=1/3$. Here, we push our calculation with bond dimension up to 60,000 and also plot the $r^{-2}$ decay as comparison, we find that the power-law decaying behavior with exponent $\eta <2$ is robust at large doping. The last observation gives evidence for quasi-long range pair-pair correlations being stabilized at these relatively large dopings. Although the short-ranged spin backgrounds appear on wider cylinders already at $\delta \sim 20\%$, pushing to higher dopings considerably strengthens the pair-pair correlations.

For $L_y=3$ cylinders, as shown in Fig.~\ref{Fig:SC} (b), the pair-pair correlations $D(r)$ decay rapidly at light doping  $\delta=1/18$.  As shown in the semi-logarithmic plot in Fig.~\ref{Fig:SC} (b), $D(r)$  could be fitted by an exponential function, meanwhile, it is  potentially also consistent with a power law $D(r) \sim 1/r^\eta$ with large exponent $\eta \sim 3.7$ for $U/t=9$ and $\eta \sim 2.8$ for $U/t=7.5$, see the double-logarithmic plot in Fig.~\ref{Fig:SC} (b). Both fitting are quite close due to the smallness of correlators both in absolute magnitude and rapid decay rate, which does not point to a robust superconducting ground state, particularly considering that $\eta > 2$ corresponds to the short-ranged correlations in 2D or the superconducting susceptibility does not diverge.  In addition, for these doping levels, we have found spatial charge modulations (see Appendix~\ref{AppedixCM}), the number of  peaks in the hole distribution function equals to the number of doped holes. This is inconsistent with a strongly paired state, where the number of peaks would be half the number of doped holes. Here, we also notice Ref.~\cite{Peng2021} fit $D(r)$ in to a power-law function with $\eta > 3.5$,  which is consistent with our results and indicates the absence of divergent susceptibility, but it was claimed to be the evidence of superconducting state~\cite{Peng2021}. At larger doping $\delta\gtrsim 20\%$ , where the SDWs are strongly suppressed [see Fig.~\ref{Fig:SDW} (a4, b4)], $D(r)$ are further enhanced.  If we  fit $D(r)$ by a power-law function , as shown in the right panel of Fig.~\ref{Fig:SC} (b),  the exponent $\eta \approx 3$ is relatively large, which suggests that  the enhancement for doping below and above $20\%$ are insignificant. This might be due to the stronger quantum fluctuations for $L_y=3$.

  \begin{table}[tbp]
 \centering
 \begin{tabular}{|c|c|c|c|}

  \hline
       {Doping} &Intermediate Coupling & Strong Coupling\\
  \hline
  \hline
   \small{Light}  & {\scriptsize{Chiral Order ($L_y=3,4$);} } & \scriptsize{SDW};\\
  $0<\delta < 10\% $  & {\scriptsize{Metallic($L_y=4$),}}&\scriptsize{Metallic}\\
  &{\scriptsize{Metallic vs. local PDW($L_y=3$)}}&\\
\hline
 \small{Moderate}  &  \scriptsize{SDW'}; & \scriptsize{SDW'};\\
  $10  \% <\delta < 20\% $  &\scriptsize{Metallic} & \scriptsize{Metallic}\\
\hline
\small{High}  &  \scriptsize{no-SDW} & \scriptsize{no-SDW} \\
   $20\% < \delta $ & \footnotesize{enhanced singlet pairing} & \footnotesize{enhanced singlet pairing}\\
    \hline
  \end{tabular}
  \caption{ Summary of results as a function of doping (light, moderate and high as defined above) and coupling strength (intermediate and strong). The SDW phase has the same wave-vector as the 120$^\circ$ magnetic order but the wave-vector in SDW' is at a potentially different ($M$) momentum point when K point is inaccessible. The anomalous chiral metal has short ranged spin correlations but long ranged spin chirality order. No clear Fermi surface is detected, unlike in the other regimes. }
  \label{Table:Summary}
 \end{table}
\begin{figure}[tbp]
\begin{center}
\includegraphics[width=0.5\textwidth]{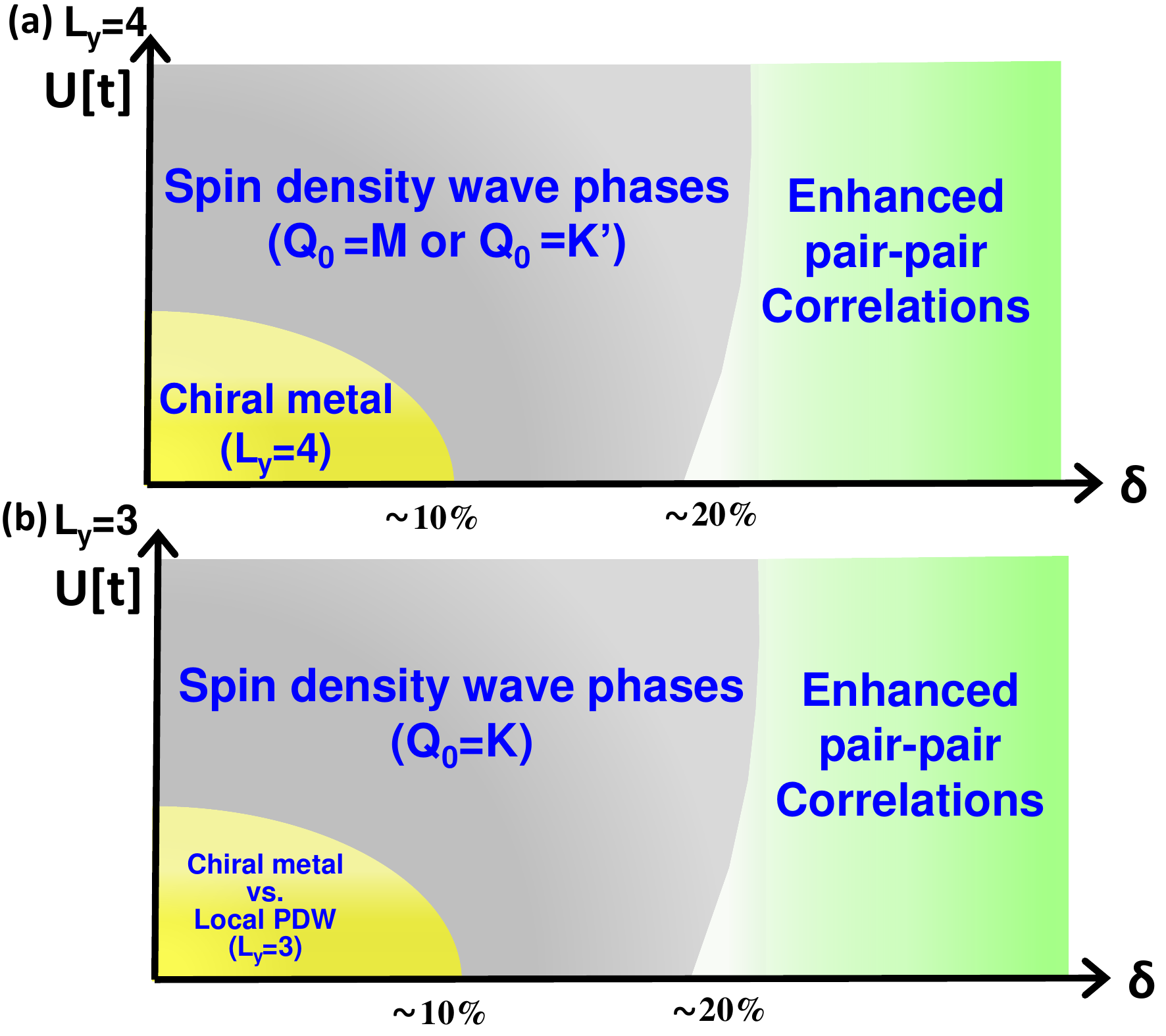}
\end{center}
\par
\renewcommand{\figurename}{Fig.}
\caption{(Color online). The schematic phase diagram of hole doped two distinct spin backgrounds  for $U/t=9$ and $U/t=18$ as a function of hole doping concentrations $\delta$. Here the chiral metallic phase is robust on $L_y=4$ cylinders while it coexists with local PDW pattern (short-range PDW) on $L_y=3$ cylinders
}
\label{Fig:PhaseDiagram}
\end{figure}

On square lattice, the previous studies mainly focused on $1/8$ doping~\cite{Ehlers2017,Jiang2019,Chung2019,Qin2019,Dolfi2015,Noack1996} and reported the exponential decay of pair-pair correlations for $L_y=4$ ladders~\cite{Ehlers2017,Jiang2019,Qin2019,Chung2019} or 2D~\cite{Qin2019}, which was attributed to the competition between charge-density waves (or stripes) and superconductivity. Here, we have identified the doping induced commensurate SDWs at $\delta\approx10\%\sim 20\%$, where the pair-pair correlations are strongly suppressed, implying the competition between SDWs and superconductivity on triangular lattice. We note that the existence of d-wave superconductivity is reported in the same triangular-lattice model for $L_y=3$ ladders at $U/t=10$~\cite{Kim2019}.
Our findings of the enhanced pairing at larger doping are consistent with hole pairing driven by spin super-exchange coupling, similar to the pairing mechanism proposed for square lattice case~\cite{PatrickLeeRMP,Dagotto1994,Zaanen2015}. An additional observation that is consistent with the superexchange scenario is that at intermediate coupling where the superexchance  $J \sim t^2/U$ exceeds that at strong coupling, the pair correlations are generally found to be stronger (compare the intermediate coupling correlators with the strong coupling correlators in the Appendix~\ref{AppedixRC}).

Here, we should point out that at doping around $\delta=1/3$, although we indeed find the pair-pair correlations are significantly enhanced than lower doping and they decay in a power-law fashion with the exponent $\eta < 2$ ($D(r) \sim 1/r^\eta$), our computations show that the single particle propagator also decays in a power-law fashion (see the Appendix~\ref{AppedixRC}), which may indicate the existence of nodal points in the superconducting phase or the superconductivity is not robust. We therefore claim it as the enhanced pair-pair correlations in the phase diagram instead of the robust superconductivity. The absence of clear superconductivity in pure Hubbard model on triangular lattice is similar to the square-lattice case as reported in Refs.~\cite{Qin2019,Jiang2019,Chung2019}. Meanwhile, we also find random sign oscillations in $D(r)$, indicating competing spatial symmetries or even more complicated nature, which we leave for future studies.

\subsection{Particle-hole Asymmetry}

Based on the above measurement, we summarize our findings in Table~\ref{Table:Summary} and set up the hole doped phase diagram as depicted in Fig.~\ref{Fig:PhaseDiagram}. Now we briefly discuss the particle-hole asymmetry with respect to electron and hole doping and leave a systematic study for future work. The particle-hole symmetry is absent on the nonbipartite triangular lattice. In contrast to the hole doped case, where light doping $\delta \sim 10\%$ does not change the spin background,
even a small density of doped electrons have a dramatic effect on the spin background  (see Appendix~\ref{AppedixED}).  For example, at intermediate coupling even for doping as low as $\delta \sim 5\%$ the maxima of the spin-spin correlation function shift to wave vector $\mathbf{M}$ (on $L_y=3$ cylinders). The distinct spin backgrounds after doping characterize the asymmetric roles of the doped holes and electrons, suggesting even richer physics on triangular lattice compared with the particle-hole symmetric square-lattice case. We leave the systematic study of electron doping and particle-hole asymmetry to future work.

\section{Conclusions}
Our study of the doped triangular lattice Hubbard model reveals that different Mott insulators obtained on changing the coupling strength $U/t$ leads to significantly different physics at  light hole doping. In the strong coupling limit a Fermi liquid with well defined hole pockets at the $\mathbf{K^\pm}$ points is observed.  In contrast at intermediate coupling, hole pockets  do not exhibit well defined quasiparticles.  Moreover long ranged spin chirality correlations
along with short ranged spin-spin correlations are observed.
These observations are consistent with spin-charge separation and spin liquid physics.
However, at  moderate doping and high doping, a SDW is established  across the range of coupling strengths and competes with superconductivity, which is established on further doping. This phenomenology should be contrasted with the competition between the emergent charge density wave (CDW) and superconductivity on the square lattice. We summarize the main features of different phases versus doping concentration $\delta$ and coupling strength $U/t$ in Table~\ref{Table:Summary} and the schematic phase diagram in Fig.~\ref{Fig:PhaseDiagram}.

Our findings for the doped Mott insulators here open up  the study of the distinct signatures of correlated electron physics  on frustrated lattices, and the inherent electron-hole asymmetry in these problems. A promising platform to experimentally study these issues is the recently realized  moir\'{e} lattice TMD or twisted TMD bilayers~\cite{TwistedTMD,TwistedTMD1}, which should be well described by the triangular lattice Hubbard model, in which the coupling strength $U/t$ is widely tunable through the twist angle, and the doping concentration is also continuously controllable. In a completely different context, in recent years, quantum simulations using the ultracold fermions in optical lattices has been significantly advanced by quantum gas miscroscopy~\cite{ColdAtom1,ColdAtom2, ColdAtom3}, which provides another platform to explore the physics in the doped Hubbard models. The experimental architecture makes it possible to tune the lattice geometry, the charge doping and the coupling strength $U/t$, and therefore, the whole phase diagram discovered in this work could be directly probed.

{\it Note added:} After this manuscript appeared, a theoretical study Ref.\cite{Song2020} also found that a  chiral metal naturally emerges on doping a Kalmeyer-Laughlin chiral spin liquid.
 
\begin{acknowledgments}
 We would like to thank Lesik Motrunich, Zheng-Yu Weng, Steven A. Kivelson, Max Metlitski, Bertrand Halperin, Zhi-Xun Shen, Ya-Hui Zhang, Ruben Verresen for valuable discussions.  ZZ acknowledges the computational resources at Harvard, CSUN and KITS, the subsequent part of this work carried out at KITS was supported by the National Natural Science Foundation of China (Grant  No. 12074375), the Fundamental Research Funds for the Central Universities (Grant  No. E0EG4303X2), the start-up funding of UCAS (Grant  No. 118900M026) and the Strategic Priority Research Program of CAS (Grant No. XDB33000000). DNS was supported by the U.S. Department of Energy, Office of Basic Energy Sciences under Grant No. DE-FG02-06ER46305 for performing larger scale DMRG simulations, AV was supported by a Simons Investigator award.
\end{acknowledgments}

\appendix

\begin{figure}[]
\begin{center}
\includegraphics[width=0.4\textwidth]{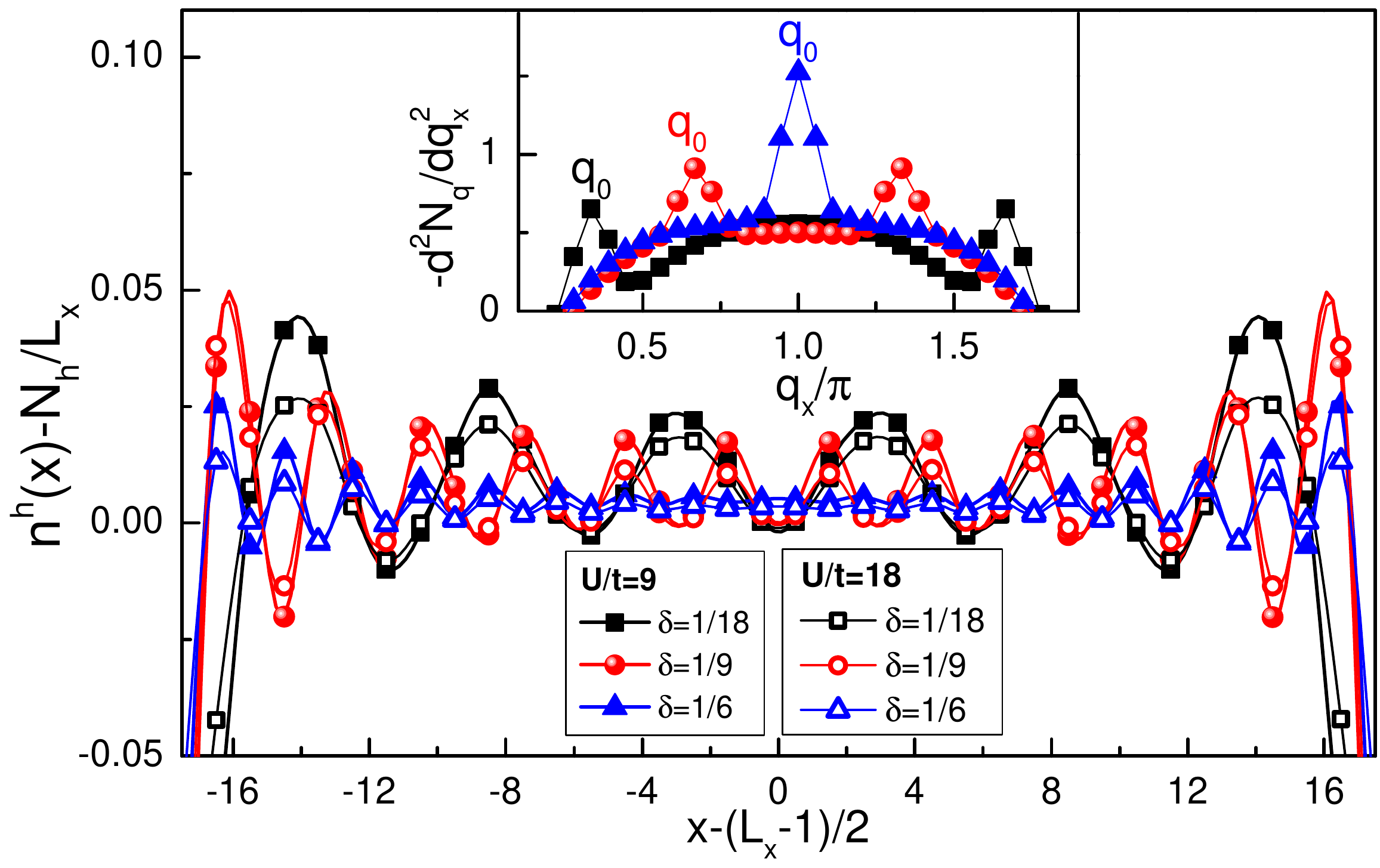}
\end{center}
\par
\renewcommand{\figurename}{Fig.}
\caption{(Color online)  The real-space hole density profile $n^h(x)$ for $U/t=9$ and $U/t=18$ at doping level $\delta=1/18, 1/9,1/6$ on $L_x=36, L_y=3$ cylinders. The inset shows the second order derivatives of the density structure factor $N_q$, where the peaks indicate the wave vectors of the charge modulations $q_0$. }
\label{Fig:Charge1}
\end{figure}
\begin{figure*}[tbp]
\begin{center}
\includegraphics[width=0.85\textwidth]{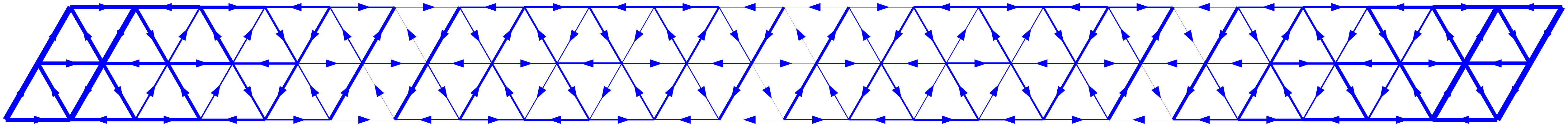}
\end{center}
\par
\renewcommand{\figurename}{Fig.}
\caption{(Color online) The charge current pattern in the chiral metal phase. This pannel shows the pattern of $I^{c}_{\langle ij\rangle}$ at $\delta=1/18$ for $U/t=9$ on $L_y=3$ cylinders, the width of each bond is proportional to the current magnitude while the arrow indicates the current direction. {Here we consider $N=24\times3$ cylinders.} The bond dimension of such calculation is set up to $12,000$.}
\label{FigS_Ly3ChargeCurrent}
\end{figure*}
\begin{figure*}[tbp]
\begin{center}
\includegraphics[width=0.85\textwidth]{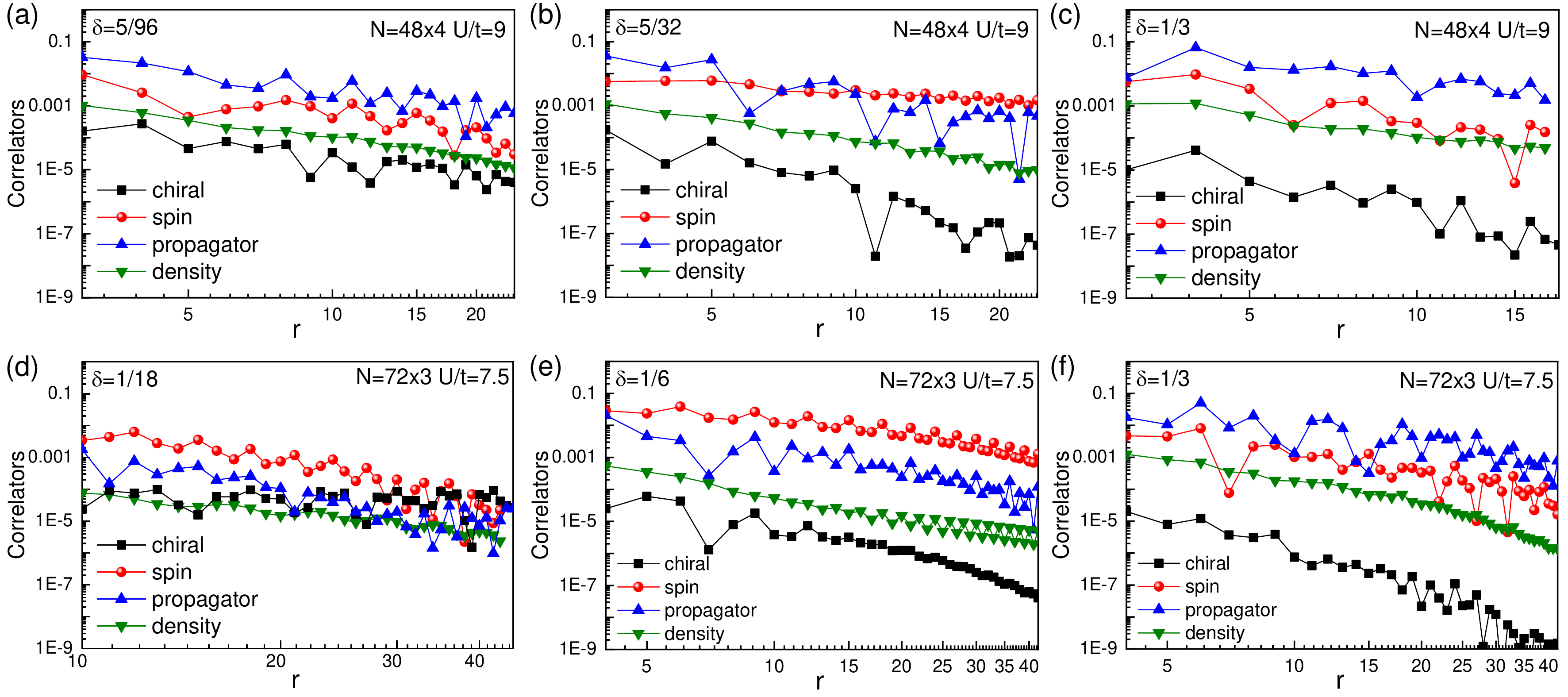}
\end{center}
\par
\renewcommand{\figurename}{Fig.}
\caption{(Color online) Real-space correlators: chiral-chiral correlations, spin-spin correlations, charge density-density correlations, single particle propagators.  Panels (a-c) and  panels (d-f) show different correlators for $U/t=9$ on $L_y=4$ and for $U/t=7.5$ on $L_y=3$ cylinders, respectively. From left to right in each row, we consider three typical hole doping concentrations: $\delta=1/18$ (a,d), $\delta=1/6$ (b,e) and $\delta=1/3$ (c,f). }
\label{FigS_Correlators}
\end{figure*}

 \section{Additional results for hole doping}
 \subsection{Charge Modulations~}\label{AppedixCM}
We first present the results of the real-space charge density distribution of the doped holes. The doped holes distribute uniformly in each rung of cylinder due to the periodical boundary conditions along $\mathbf{e_y}$,  we thus focus on the hole distribution along $\mathbf{e_x}$ and define ${n^h}\left( x \right) \equiv \sum_{y = 1}^{{L_y}} {{n^h}\left( {x,y} \right)}$, where $n^h(x,y)\equiv 1-n(x,y)$ denotes the hole density on the site with coordinate $(x,y)$. As shown in Fig.~\ref{Fig:Charge1} for $U/t=9$ and  $U/t=18$, the doped holes exhibit strongly spatial modulations  at  $\delta\lesssim 20\%$ for both cases, while their amplitude decrease with the increase of hole concentration. The wave vectors of the charge modulations $q_0$ can be determined by the singularity/kinks in the structure factors $N_q$, as illustrated in the second order derivative of $N_q$  in the inset of Fig.~\ref{Fig:Charge1}. Moreover, the number of peaks in $n^h(x)$ equals to the number of doped holes, implying the absence of strongly pairing state.

\subsection{Charge current pattern for $L_y=3$ cylinders}\label{AppedixCC}

In the main text, we have shown the the bond charge current pattern for the chiral metallic phase on $L_y=4$ cylinders. Fig.~\ref{FigS_Ly3ChargeCurrent} shows the bond charge current pattern on $L_y=3$ cylinders with $U/t=9$, the width of each bond is proportional to the current magnitude and the arrow indicates the current direction, the bond current pattern exhibits the translational symmetry breaking in the bulk along the horizontal direction, though we also notice the existence of domains in the pattern. Meanwhile, the time-reversal invariant spin current $I^{s}_{ij}\equiv -\text{i}\langle{S}_{i}^{+}{S}_{j}^{-}-S_{i}^{-}{S}_{j}^{+}\rangle$ is vanishingly small in the same phase, in consistent with the existence of the spin chirality order for chiral metal.

\begin{figure}[tbp]
\begin{center}
\includegraphics[width=0.45\textwidth]{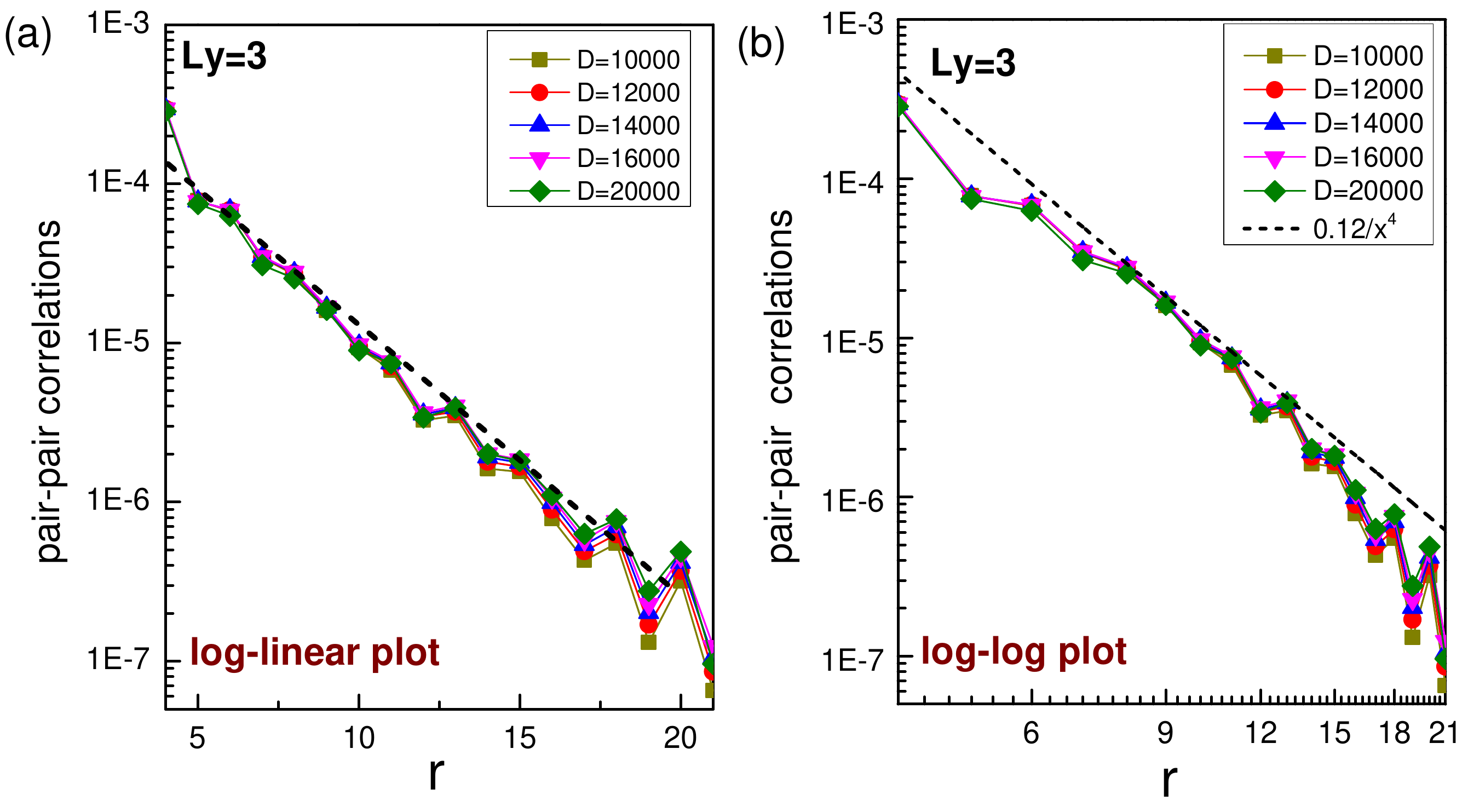}
\end{center}
\par
\renewcommand{\figurename}{Fig.}
\caption{(Color online)The pair-pair correlations for { $N=36\times3$ cylinders} at doping $\delta=1/18$.  Panel (a) shows the plots in semi-logarithmic scale, while panel (b) shows the same data but in double-logarithmic scale. }
\label{Fig:3LegSCD}
\end{figure}
\begin{figure}[tbp]
\begin{center}
\includegraphics[width=0.45\textwidth]{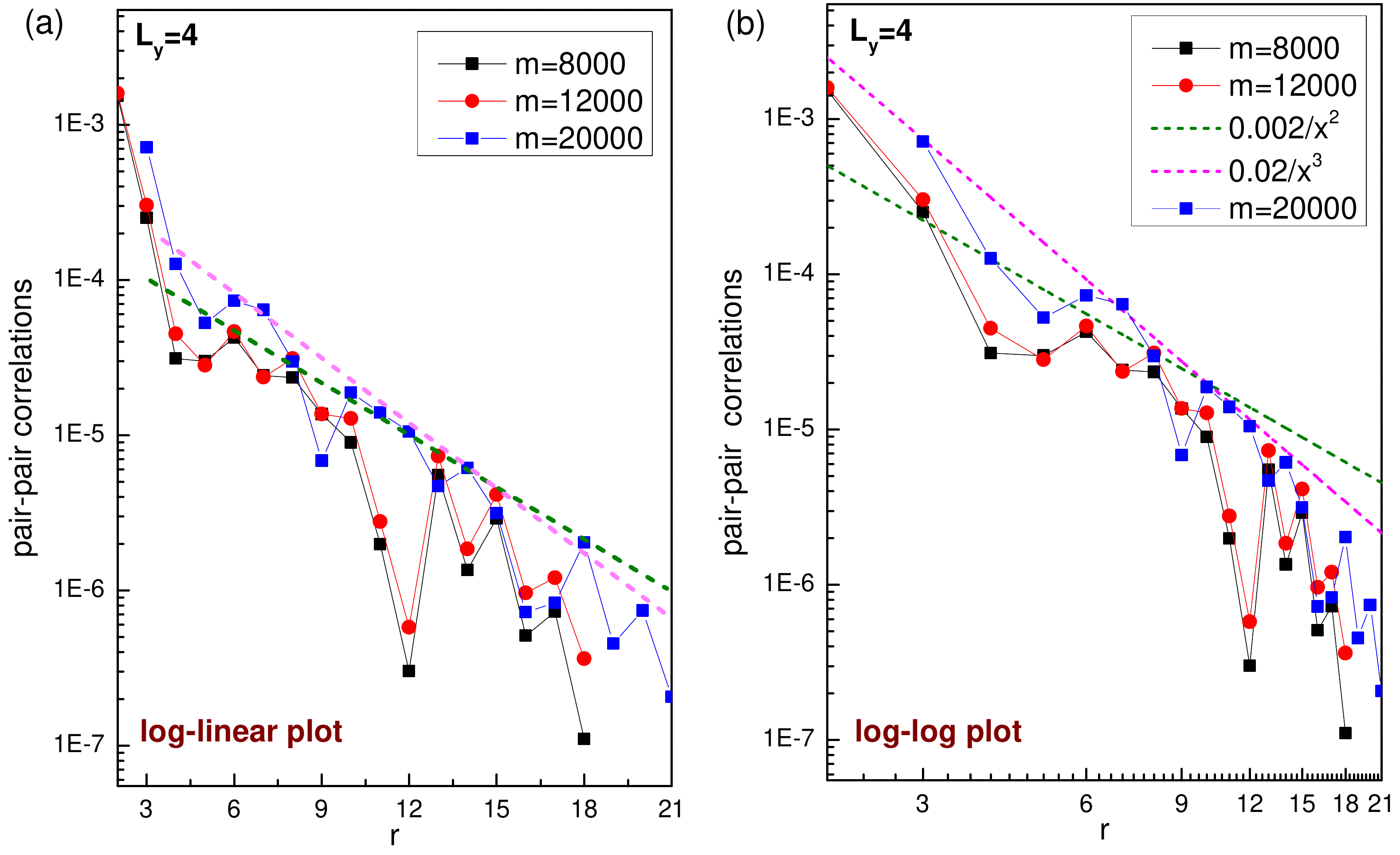}
\end{center}
\par
\renewcommand{\figurename}{Fig.}
\caption{(Color online)The pair-pair correlations for {$N=30\times4$ cylinders} at doping $\delta=1/20$.  Panel (a) shows the plots in semi-logarithmic scale, while panel (b) shows the same data but in double-logarithmic scale.}
\label{Fig:4LegSCD}
\end{figure}
\begin{figure}[tbp]
\begin{center}
\includegraphics[width=0.5\textwidth]{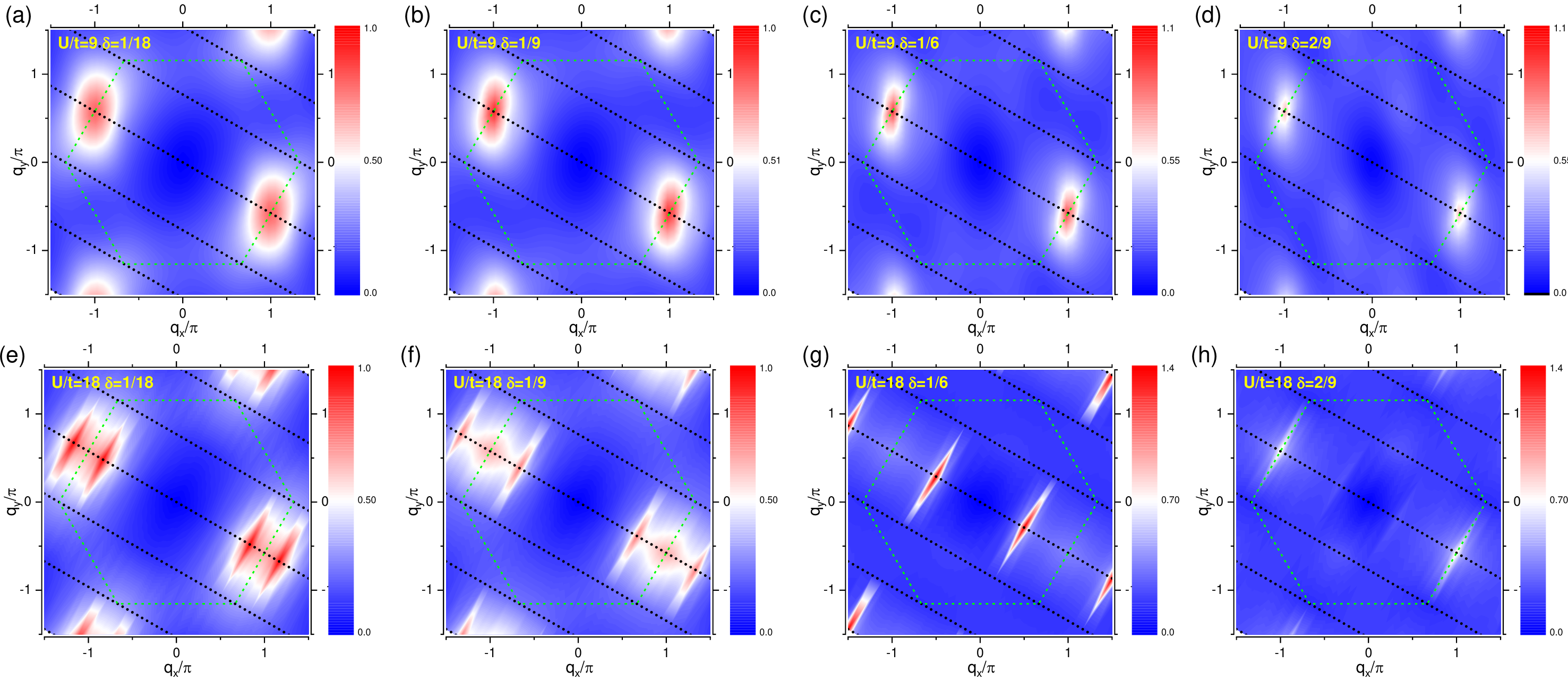}
\end{center}
\par
\renewcommand{\figurename}{Fig.}
\caption{(Color online) The contour plot of spin structure factor $S_\mathbf{q}$  for $U/t=9$ [(a) to (d)] and $U/t=18$ [(e) to (h)] with different electron doping concentrations $\delta$. From left to right in each row, the concentrations of the doped electrons are $\delta=1/18$ (a, e), $\delta=1/9$ (b,f), $\delta=1/6$ (c, g), $\delta=2/9$ (d, h). The black dots represent the momentum points we can access in the  Brillouin zone (dashed line) of $N=36\times3$ cylinders.  }
\label{Fig:SDW-Electron}
\end{figure}

\subsection{Various real-space correlators}\label{AppedixRC}

To further confirm the dominant physics in each phase, we compute various correlation functions in the real space and put them into the same figure to compare, including: (i) the spin-spin correlation: $\left\langle {S_i^zS_{i+r}^z} \right\rangle$; (ii) the charge density-density correlation:  $\left\langle {n_i n_{i+r}} \right\rangle- \langle n_i\rangle \langle n_{i+r}\rangle$; (iii) the single particle propagator: $\sum_{\sigma }\langle c_{i\sigma }^\dag c_{i+r, \sigma }\rangle $; and (iv) the chiral-chiral correlations $|\langle C_{\bigtriangleup_i} C_{\bigtriangleup_{i+r}}\rangle|$, as shown in Fig.~\ref{FigS_Correlators}.  Figures~\ref{FigS_Correlators} (a-c) and (d-f) show $L_y=4, L_x=48$ cylinders and  $L_y=3, L_x=72$ cylinders at three typical dopings, respectively.

\subsection{Pair-Pair correlations}

We next examine the  pair-pair correlations $D(r)$ at hole doping side. In the main text, we have shown the results on longer cylinders with size $N=72\times3$ and $N=48\times4$. In the appendix, we show the bond-dimension dependence of the pair-pair correlations.

At light doping, the pairing correlators appear to decay exponentially and there is no systematic change of correlators on increasing bond dimension on small clusters with $N=36\times3$, as shown in Fig.~\ref{Fig:3LegSCD} for $L_y=3$ cylinders and  Fig.~\ref{Fig:4LegSCD} for $L_y=4$ cylinders. If we fit $D(r)$ with a power law function $D(r)\sim r^{-\alpha}$ as best we can, the resulting exponent is relatively large. For instance, $\alpha\gtrsim4$  for $\delta=1/18$ when $L_y=3$ [see Fig.~\ref{Fig:3LegSCD} (b)] and $\alpha\gtrsim3$  for $\delta=1/20$ when $L_y=4$ [see Fig.~\ref{Fig:4LegSCD} (b)]. Furthermore,  while this behavior is potentially also consistent with a power law with large exponent, we note that the smallness of correlators both in absolute magnitude and rapid decay rate does not point to a superconducting ground state at small and medium doping. In addition, for these doping levels, we have found spatial charge modulations  in Fig.~\ref{Fig:Charge1}, the number of  peaks in the hole distribution function equals to the number of doped holes. This is inconsistent with a strongly paired state, where the number of peaks would be half the number of doped holes.

\section{Electron Doping}\label{AppedixED}
In the main text, we mainly focus on the hole doped side, in this section, we very briefly compare with electron doping side.  For the electron doping,  we choose $U/t=9$ and $U/t=18$ for comparative study on $L_y=3$ cylinders at doping $\delta>5\%$. Here, we should point out that the comprehensive study of electron doping lies outside of the scope of the present work, which we will leave for future systematical investigations. 

We examine the spin channel by measuring the static spin structure factor $S_\mathbf{q} (\mathbf{Q}) =\frac{1}{N}\sum_{i,j} {\left\langle {S_i^zS_j^z} \right\rangle e^{i \mathbf{Q}\cdot(\mathbf{r_i}-\mathbf{r_j})}}$.
 Figures~\ref{Fig:SDW-Electron} (a-d) and (e-h) show the contour plot of $S_\mathbf{q}$ against electron doping with $U/t=9$  and $U/t=18$, respectively. For the electron doping with $U/t=9$, $S_\mathbf{q}$ exhibits peaks at commensurate momentum $\mathbf{q=M}$  up to around $\delta=1/6$ [see Fig.~\ref{Fig:SDW-Electron} (a-c)], indicating the commensurate SDWs.  For the electron doped  120$^{\circ}$ N\'eel ordered spin background with $U/t=18$ [see Fig.~\ref{Fig:SDW-Electron} (e-g)], $S_\mathbf{q}$ exhibits splitting peaks around $\mathbf{q=M}$, suggesting incommensurate SDWs. At larger doping $\delta > 20\%$, the spin backgrounds are indistinguishable for $U/t=9$ and  $U/t=18$ [see Fig.~\ref{Fig:SDW-Electron} (d) and (h)], the spin correlations become short ranged for both coupling strength. The different responses of the spin backgrounds against doping can be served as an evidence of the asymmetry with respect to the electron and hole doping on triangular lattice.

\end{document}